\documentclass[%
 aip, jcp,
 amsmath,amssymb,
reprint,%
floatfix,
]{revtex4-1}

\usepackage{graphicx}
\usepackage{dcolumn}
\usepackage{bm}
\usepackage{multirow}
\usepackage{csquotes}
\usepackage{algpseudocode}
\usepackage{amsmath}
\usepackage{xcolor}

\newcommand\bra[1] {\langle {#1} |}
\newcommand\ket[1] {| {#1} \rangle}
\newcommand\braket[2] {\langle {#1} | {#2} \rangle}

\newcommand{\themi}{M_I}

\newcommand{\thezi}{Z_I}
\newcommand{\thepi}{\mathbf{P}_I}

\newcommand{\thepii}{\boldsymbol{\Pi}_I}

\newcommand{\thechii}{\boldsymbol{\chi}_I}
\newcommand{\theri}{\mathbf{R}_I}

\newcommand{\theai}{\mathbf{A}(\mathbf{R}_I)}

\newcommand{\theb}{\mathbf{B}}

\newcommand{\thennuc}{N_\mathrm{nuc}}

\newcommand{\theoph}{H}
\newcommand{\theophel}{H_\mathrm{el}}
\newcommand{\theoptnuc}{T_\mathrm{nuc}}

\newcommand{\thewf}{\Psi (\mathbf{R}, \mathbf{r}, t)}
\newcommand{\thewfel}{\phi}
\newcommand{\thewfnuc}{\psi}

\newcommand{\thewfelfull}{\phi(\mathbf{r}; \mathbf{R})}
\newcommand{\thewfnucfull}{\psi (\mathbf{R}, t)}

\newcommand{\theebo}{U_\mathrm{BO}}

\newcommand{\theG}{\mathbf{G}}
\newcommand{\thenel}{N_\mathrm{el}}
\newcommand{\thee}{e}
\newcommand{\theari}{\mathbf{A}(\mathbf{r}_i)}
\newcommand{\thepiel}{\mathbf{p}_i}

\begin{document}

\title{\emph{Ab-Initio} Molecular Dynamics with Screened Lorentz Forces. Part I. Calculation and Atomic Charge Interpretation of Berry Curvature
}

\author{Tanner Culpitt}
\affiliation
{Hylleraas Centre for Quantum Molecular Sciences,  Department of Chemistry, 
University of Oslo, P.O. Box 1033 Blindern, N-0315 Oslo, Norway}
\author{Laurens D. M. Peters}
\affiliation
{Hylleraas Centre for Quantum Molecular Sciences,  Department of Chemistry, 
University of Oslo, P.O. Box 1033 Blindern, N-0315 Oslo, Norway}
\author{Erik I. Tellgren}
\email{erik.tellgren@kjemi.uio.no}
\affiliation
{Hylleraas Centre for Quantum Molecular Sciences,  Department of Chemistry, 
University of Oslo, P.O. Box 1033 Blindern, N-0315 Oslo, Norway}
\author{Trygve Helgaker}
\affiliation
{Hylleraas Centre for Quantum Molecular Sciences,  Department of Chemistry, 
University of Oslo, P.O. Box 1033 Blindern, N-0315 Oslo, Norway}

\date{\today}

\begin{abstract}
The dynamics of a molecule in a magnetic field is significantly different form its zero-field counterpart. One important difference in the presence of a field is the Lorentz force acting on the nuclei, which can be decomposed as the sum of the bare nuclear Lorentz force and a screening force due to the electrons. This screening force is calculated from the Berry curvature and can change the dynamics qualitatively. It is therefore important to include the contributions from the Berry curvature in molecular dynamics simulations in a magnetic field. In this work, we present a scheme for calculating the Berry curvature numerically, by a finite-difference technique, addressing challenges related to the arbitrary global phase of the wave function. The Berry curvature is calculated as a function of bond distance for H$_2$ at  the restricted and unrestricted Hartree--Fock levels of theory and for CH$^{+}$ as a function of the magnetic field strength at the restricted Hartree--Fock level of theory. The calculations are carried out using basis sets of contracted Gaussian functions equipped with London phase factors (London orbitals) to ensure gauge-origin invariance. In the paper, we also interpret the Berry curvature in terms of atomic charges and discuss its convergence in basis sets with and without London phase factors. Calculation of the Berry curvature allows for its inclusion in \textit{ab initio} molecular dynamics simulations in a magnetic field.
\end{abstract}

\maketitle

\section{Introduction}

In quantum chemistry and quantum mechanics more generally, the Berry phase\cite{Berry1984,Resta2000,Zwanziger1990,Moore1991,Mead1992} and Berry curvature\cite{Resta2000} are prevalent and important concepts, particularly when dealing with molecular dynamics in magnetic fields. For the purposes of this work, we are interested in how the Berry curvature manifests itself in molecular dynamics equations. We concern ourselves here with Born--Oppenheimer molecular dynamics (BOMD),\cite{marx2009ab,Carvalho2014} which is predicated upon the Born--Oppenheimer approximation\cite{born1927} where the electrons of an atomic or molecular system are assumed to respond instantaneously to any perturbations in nuclear coordinates. The total approximate ground-state wave function is thereby written as a product of the electronic and nuclear wave functions, although in principle the exact wave function can also be represented in product form.\cite{HUNTER_IJQC9_237,Abedi2010,Requist2016} When formulating BOMD in the absence of fields, where the electronic wave function can be taken as real, the nuclear Hamiltonian can be approximated by the sum of the nuclear kinetic energy and a scalar potential. In this setting, the equations of motion for the nuclei reduce to Newton's equations, with the conservative force on each nucleus  calculated as the negative gradient of the electronic potential-energy surface.

For BOMD in a magnetic field, the situation is more complicated because of the velocity-dependent forces. The electronic wave function is generally complex and the nuclear Born--Oppenheimer Hamiltonian contains contributions from both magnetic and geometric vector potentials. The magnetic vector potential gives rise to the bare Lorentz force acting on the nuclei, while the geometric vector potential gives rise to the Berry force, representing the effect of the screening of the magnetic field by the electrons in the system.\cite{Resta2000,Ceresoli2007,Schmelcher1988,Detmer1995} The   total effective Lorentz force acting on the nuclei is the sum of the bare Lorentz force and the Berry screening force. The screening force can be large and change the dynamics qualitatively, making its implementation an important step towards a full realization of BOMD in a magnetic field, as described in a companion article to the present paper.

The main challenge associated with calculating the Berry curvature from finite-difference is that the global phase of the wave function for each nuclear perturbation is in principle arbitrary. It is therefore important to have a consistent, well-justified way of dealing with the phase. Previously, Ceresoli, Marchetti, and Tosatti have obtained the Berry force from the Berry curvature for H$_2$ using a simple model electronic wave function.\cite{Ceresoli2007} Here, we present a general finite-difference implementation of the Berry curvature within the software package {\sc london} for \emph{ab initio} molecular electronic-structure calculations in a finite magnetic field.\cite{LondonProgram} The {\sc London} program uses London atomic orbitals (also known as gauge-including atomic orbitals (GIAOs)))\cite{London1937,Hameka1958,Ditchfield1976,Helgaker1991,Tellgren2008,Tellgren2012,Irons2017,Pausch2020} for gauge-origin invariant calculation of energies and molecular properties at various levels of theory including Hartree--Fock theory,\cite{Tellgren2008,Tellgren2009,Tellgren2012} (current-)density-functional theory,\cite{TELLGREN_JCP140_034101,FURNESS_JCTC11_4169}  full-configuration-interaction theory,\cite{Lange2012,Austad2020} coupled-cluster theory,\cite{Stopkowicz2015} and linear-response theory.\cite{Sen2019}

This work is organized as follows. Section II contains a derivation of the effective nuclear Hamiltonian in the presence of a magnetic field, along with the resulting equations of motion and the finite difference scheme for calculation of the Berry curvature. Justification for the phase corrections used to account for the arbitrary phase of the electronic wave function is also presented. Section III presents Berry curvature results for H$_2$ and CH$^+$ at the Hartree--Fock level of theory.
For H$_2$, comparisons are made with the previous results of Ref.\,\onlinecite{Ceresoli2007}. The work is summarized and future directions are given in Section~IV.

\section{Theory}

We consider a joint system of nuclei and electrons. Throughout this work, $I$ and $J$  serve as indices for the $\thennuc$ nuclei, and their lowercase counterparts $i$ and $j$ will serve as indices for the $\thenel$ electrons. We use the notation $\themi$, $\thezi$, and $\theri$, and for the  mass, atom number, and position of nucleus $I$, respectively.  We use $\mathbf{r}_i$ and $\mathbf{p}_i$ for the position operator and momentum operator of electron $i$, respectively. 
The vectors of collective nuclear and electronic coordinates are denoted by $\mathbf{R}$ and $\mathbf r$, respectively.
The vector potential of a uniform magnetic field $\mathbf B$ at position $\mathbf u$ is given by $\mathbf A(\mathbf u) = \frac{1}{2} \theb \times (\mathbf u-\theG)$, where $\theG$ is the gauge origin. 

\subsection{Born--Oppenheimer approximation}

We are interested in writing down the effective nuclear Hamiltonian within the Born--Oppenheimer approximation, beginning from the nonrelativistic Hamiltonian of a molecular system in a uniform magnetic field 
\begin{equation}
\theoph_\text{mol} = \theoptnuc + \theophel + V_\text{nuc}\ .
\label{ham_000}
\end{equation}
Here the nuclear kinetic energy operator is given by
\begin{equation}
\theoptnuc = \sum \limits_{I=1}^{\thennuc} \dfrac{\Pi_I^2}{2\themi} , \quad\thepii = \thepi - \thezi e \theai \ ,
\label{ham_001}
\end{equation}
where $\thepi = - \mathrm i\hbar \partial/\partial \theri$ is the canonical
momentum and $\thepii$ the physical momentum of nucleus $I$, while the
nuclear repulsion operator is given by
\begin{equation}
V_\text{nuc} = \sum_{I>J=1}^{N_\text{nuc}}\frac{Z_I Z_J e^2}{4 \pi \varepsilon_0\vert \mathbf R_I - \mathbf R_J\vert}
\end{equation}
where $e$ is the elementary charge and $\varepsilon_0$ the vacuum permittivity. The  electronic Hamiltonian is given by
\begin{align}
\theophel &= \frac{1}{2m_\text{e}} \sum_{i=1}^{\thenel} (\thepiel + \thee \theari)^2 
 \nonumber \\ &
 + \sum_{i>j=1}^{N_\text{el}}\frac{e^2}{4 \pi \varepsilon_0\vert \mathbf r_i - \mathbf r_j\vert}  - \sum_{i=1}^{N_\text{el}}\sum_{I=1}^{N_\text{nuc}}\frac{Z_I e^2}{4 \pi \varepsilon_0\vert \mathbf r_i - \mathbf R_I\vert}
\label{v_nn} 
\end{align}
The  molecular wave function satisfies the time-dependent Schr\"odinger equation
\begin{equation}
H_\text{mol} \ket{\Psi} = \mathrm{i} \hbar \frac{\partial }{\partial t} \ket{\Psi} .
\label{molsch}
\end{equation}
Staying within the Born--Oppenheimer approximation, we write the molecular
wave function as a product of a time-dependent nuclear wave function 
$\psi(\mathbf R,t)$ and an electronic
wave function $\phi(\mathbf{r};\mathbf{R})$ that is time independent:
\begin{equation}
\thewf = \thewfnucfull \thewfelfull.
\label{wf_bo}
\end{equation}
The relevant configurations $\mathbf{R}$ where the nuclear wave function is concentrated evolves in time, but the electronic wave function formally does not.
We assume that the electronic wave function is obtained at each nuclear configuration $\mathbf{R}$ by solving the time-independent electronic Schr\"odinger equation in some exact or approximate manner such that
\begin{equation}
   U_\text{BO}(\mathbf R) = \bra{\phi}  H_\text{el} \ket{\phi}
   \label{UBO}
\end{equation}
and wish to solve the molecular Schr\"odinger Eq.\,\eqref{molsch} approximately by
inserting the product form of Eq.\,\eqref{wf_bo} and projecting the resulting equation from the left by the electronic wave function,
\begin{equation}
\bra{\phi}  H_\text{mol} \ket{\psi \phi} = \mathrm{i} \hbar \bra{\phi}  \frac{\partial }{\partial t} \ket{\psi \phi}.
\end{equation}
By constructing an effective
nuclear Hamiltonian $H$ such that
\begin{align}
H \vert \psi \rangle = \bra{\thewfel} T_\text{nuc} + H_\text{el} \ket{\thewfnuc \thewfel} \, , 
\label{ham_004}
\end{align}
we may determine the  Born--Oppenheimer nuclear wave function from the effective nuclear Schr\"odinger equation
\begin{equation}
    H \vert \psi \rangle = \mathrm i \hbar \frac{\partial}{\partial t} \vert \psi \rangle \, .
\label{nucTDSE}
\end{equation}
Within the Born--Oppenheimer approximation, all time dependence of the molecular system is determined by the nuclear Schr\"odinger Eq.\,\eqref{nucTDSE}. Loosely speaking, the electronic state adjusts instantaneously to the motion of the nuclei. 

\subsection{Geometric vector and scalar potentials}

To determine the Born--Oppenheimer nuclear Hamiltonian $H$ from Eq.~\eqref{ham_004}, we consider first the projection of the squared kinetic momentum of nucleus $I$ on the electronic wave function. Projecting against the electronic state, we obtain after some algebra
\begin{equation}
    \bra{\phi} \Pi_I^2 \ket{\psi \phi} =
    \left(  \Pi_I^2 + 2 \boldsymbol \chi_I \cdot \thepii + \Lambda\right) \ket{\psi},
    \label{Pisq}
\end{equation}
where we have introduced 
the geometric vector and scalar potentials, respectively:
\begin{align}
\thechii &= \bra{\phi} \mathbf{P}_I \ket{\phi},
\label{gvp_def} \\
\Lambda_I &= \langle \phi \vert P_I^2 \vert \phi \rangle \, .
\label{lambda_def}
\end{align}
The geometric vector potential is also known as the Berry potential or Berry connection. Application of the momentum operator $\mathbf P_I = - \mathrm{i} \hbar \boldsymbol \nabla_I$ to $\braket{\phi}{\phi} \equiv 1$ gives
\begin{equation}
    \mathbf{P}_I \braket{\phi}{\phi} = - \braket{\mathbf{P}_I \phi}{\phi} + \braket{\phi}{\mathbf{P}_I \phi} = \mathbf{0}.
\end{equation}
Hence, it follows that the geometric vector potential is real-valued,
\begin{equation}
    \boldsymbol{\chi}_I = \braket{\phi}{\mathbf{P}_I \phi} = \braket{\mathbf{P}_I \phi}{\phi} = \braket{\phi}{\mathbf{P}_I \phi}^\ast = \boldsymbol \chi_I^\ast.
\end{equation}
Next, we observe that
\begin{equation}
\mathbf{P}_I \cdot \boldsymbol{\chi}_I = - \braket{\mathbf{P}_I \phi}{\mathbf{P}_I \phi} + \braket{\phi}{P_I^2 \phi}.
\end{equation}
Introducing
\begin{align}
    \Delta_I = \braket{\mathbf{P}_I \phi}{\mathbf{P}_I \phi} = \Delta_I^\ast, \label{DeltaI}
\end{align}
we find that the  scalar potential may be decomposed into real and imaginary parts in the manner
\begin{equation}
    \Lambda_I = \Delta_I + \mathbf P_I \cdot \boldsymbol \chi_I \,.
\end{equation}
To express Eq.\,\eqref{Pisq} in a more convenient form, we introduce an effective nuclear physical momentum
\begin{align}
    \overline {\boldsymbol \Pi}_I &= \boldsymbol \Pi_I + \boldsymbol \chi_I
    \nonumber \\ &= \mathbf P_I - Z_I e \mathbf A(\mathbf R_I) + \boldsymbol \chi_I
    \label{eff_mom}
\end{align}
and the diagonal Born--Oppenheimer correction (DBOC)
\begin{align}
    \overline \Delta _I &= \Delta_I - \chi_I^2,
\label{dboc_bar}
\end{align}
and obtain the following projected squared momentum operator
\begin{equation}
    \bra{\phi} \Pi_I^2 \ket{\psi \phi} =
    \left( \overline{\Pi}_I^2 + \overline{\Delta}_I\right) \ket{\psi}.
    \label{Pisq2}
\end{equation}
By introducing the resolution of identity in Eq.~\eqref{DeltaI}, we find that the DBOC is non-negative and may be written as
\begin{align}
    \overline{\Delta}_I  
   &= \sum_{p\neq 0} \big\vert \bra{\phi_p} \mathbf{P}_I \ket{\phi} \big\vert^2 \geq 0,
   \label{odeltai}
\end{align}
where the summation is over all excited states. 

\subsection{Born--Oppenheimer nuclear Schr\"odinger equation}

Returning to the Schr\"odinger equation in Eq.\,\eqref{nucTDSE},
we find that it may be written in the form
\begin{align}
    H \ket{\psi} = (T + U) \ket{\psi} = 
    \mathrm{i} \hbar \frac{\partial}{\partial t} \ket{\psi}
    \label{HTUop}
\end{align}
in terms of the kinetic- and potential-energy operators
\begin{align}
T &=\sum_{I=1}^{N_\text{nuc}}\frac{1}{2M_I}\overline \Pi_I^2 \,,
\label{Top}
\\
U &=  U_\text{BO} + \sum_{I=1}^{N_\text{nuc}}\frac{1}{2M_I} \overline \Delta_I + V_\text{nuc}\,.
\label{Uop}
\end{align}
We note that all quantities entering the Hamiltonian depend on the nuclear coordinates $\mathbf R$, including the momentum operators.

The mass-dependent DBOC contribution to the potential is normally neglected but has in a few cases been calculated as a nonadiabatic correction to the potential-energy surface.\cite{Handy1986,Ioannou1996,Valeev2003} In  these cases, the wave
function has been real so that the geometric vector potential vanishes and $\overline \Delta_I = \Delta_I$. The correction is expected to be small and have a negligible impact on molecular dynamics in regions where the Born--Oppenheimer approximation is valid. However, it is in general more important to include than in the field-free case, especially for strong fields and/or dynamics far from equilibrium.\cite{Schmelcher1988} We finally note that the Born--Oppenheimer approximation is usually taken to mean that all matrix elements that couple different electronic states are neglected; since the DBOC arises from such couplings (Eq.\,\eqref{odeltai}), its inclusion is normally thought of as going beyond the Born--Oppenheimer approximation.

\subsection{Geometric gauge transformations}

The effective nuclear Hamiltonian $H = T + U$ given in Eqs.\,\eqref{HTUop}--\eqref{Uop} depends on several expectation values over the geometry-dependent electronic wave function---namely, the Born--Oppenheimer potential $U_\text{BO}(\mathbf R)$, the geometric vector potential $\boldsymbol \chi_I(\mathbf R)$, and the geometric scalar potential $\overline \Delta_I(\mathbf R)$, whose geometry dependence arises from the parametric  dependence of the electronic wave function on the nuclear coordinates. Since the electronic wave function at each geometry is determined by solving the electronic Schr\"odinger equation, it is (in the absence of degeneracies) uniquely determined up to a geometry-dependent phase factor. It is therefore important to establish the invariance of the nuclear Schr\"odinger equation to a geometry-dependent gauge transformation of the electronic wave function.

Consider the following geometry-dependent gauge transformation of the electronic wave function
 \begin{equation}
 \ket{\thewfel'}  = \mathrm{e}^{-\mathrm{i}  F(\mathbf{R})/\hbar} \ket{\thewfel} \,,
\end{equation}
where the gauge function $F$ is a real-valued differentiable function of the nuclear coordinates $\mathbf{R}$.
The corresponding gauge-transformed matrix elements entering the nuclear Hamiltonian $H = T + U$ are
\begin{align}
    U_\text{BO}^\prime(\mathbf R) &= U_\text{BO}(\mathbf R) \,,
    \\
    \boldsymbol \chi_I^\prime(\mathbf R) &= \boldsymbol \chi_I(\mathbf R) - \boldsymbol \nabla_I F(\mathbf R) \,, \label{chigt}
    \\
    \overline \Delta_I^\prime(\mathbf R) &=   \overline \Delta_I(\mathbf R) \,.
    \label{deltagt}
\end{align}
The  invariance of $U_\text{BO}(\mathbf R)$ follows 
since $H_\text{el}$ in Eq.\,\eqref{UBO} does not differentiate with respect to the nuclear coordinates. To determine the gauge transformations of $\boldsymbol \chi_I(\mathbf R)$ in Eq.\,\eqref{gvp_def} and of $\overline \Delta_I(\mathbf R)$ in Eq.\,\eqref{odeltai}, we note that
\begin{equation}
\mathbf{P}_I \ket{\phi^\prime} = e^{-\mathrm i F(\mathbf R)/\hbar}( \mathbf{P}_I \ket{\phi} - \ket{\phi} \boldsymbol{\nabla}_I F(\mathbf{R})),
\end{equation}
which upon projection by $\bra{\phi^\prime}$ and $\bra{\phi^{\prime}_p}$, respectively, followed by use of orthonormality $\braket{\phi}{\phi} = 1$ and $\braket{\phi_p}{\phi} = 0$ gives Eqs.~\eqref{chigt} and~\eqref{deltagt}.

Using these results, we find that the nuclear momentum operator transforms as
\begin{align}
\overline \Pi_I^\prime &= \overline \Pi_I - \boldsymbol \nabla_I F(\mathbf R)   \nonumber  \\
&= \mathrm e^{\mathrm i F(\mathbf R)/\hbar} \,\overline \Pi_I\, \mathrm e^{-\mathrm i F(\mathbf R)/\hbar}
\end{align}
and hence that the operator of the time-dependent Born--Oppenheimer Schr\"odinger equation transforms as
\begin{equation}
    H^\prime - \mathrm i \hbar \frac{\partial}{\partial t} = \mathrm e^{\mathrm i F(\mathbf R)/\hbar} \left( H - \mathrm i \hbar \frac{\partial}{\partial t}
    \right)\mathrm e^{-\mathrm i F(\mathbf R)/\hbar} \,,
\end{equation}
whether or not the DBOC is included in the Hamiltonian. This gauge transformation may be compensated for by a unitary transformation of the nuclear wave function
\begin{equation}
\ket{\psi^\prime} = \mathrm{e}^{\mathrm{i} F(\mathbf{R})/\hbar} \ket{\psi}
\end{equation}
and hence does not affect any observable quantities if the variational space is sufficiently flexible to accomodate such transformations.

In passing, we note that, while a geometry-dependent gauge transformation can change the local behavior of the Berry connection $\thechii(\mathbf{R})$, there is an invariant phase along any closed loop $C$,
\begin{equation}
\zeta = \oint_\mathrm C \thechii(\mathbf R) \cdot d\mathbf{R},
\label{eom_t018}
\end{equation}
known as the Berry phase.~\cite{Berry1984} 

\subsection{Magnetic gauge transformations}

Consider now a gauge transformation of the external magnetic vector potential
\begin{equation}
\tilde{\mathbf{A}}(\mathbf u) = \mathbf A(\mathbf u) + \boldsymbol \nabla f(\mathbf u)
\end{equation}
and the corresponding gauge-transformed nuclear and electronic wave functions
\begin{align}
    \ket{\tilde{\phi}} &= \left(\prod_{i=1}^{N_\text{el}} \mathrm{e}^{-\mathrm{i} e f(\mathbf r_i)/\hbar} \right) \ket{\phi}, \\
    \ket{\tilde{\psi}} &= \left(\prod_{I=1}^{N_\text{nuc}} \mathrm{e}^{\mathrm{i} Z_I e f(\mathbf R_I)/\hbar}\right)  \ket{\psi}.
    \label{nucwf_gauge_trans}
\end{align}
In this case, all electronic expectation values entering the nuclear Hamiltonian, including the geometric vector potential, are unaffected
\begin{align}
   \tilde  U_\text{BO}(\mathbf R) &= U_\text{BO}(\mathbf R)\,,
    \\
    \tilde{\boldsymbol \chi}_I(\mathbf R) &= \boldsymbol \chi_I(\mathbf R)\,,
    \\
   \tilde{ \overline \Delta}_I(\mathbf R) &=   \overline \Delta_I(\mathbf R) \,,
\end{align}
while the nuclear momenta transform as
\begin{align}
\tilde{\overline \Pi}_I &= \overline \Pi_I - Z_I e \boldsymbol \nabla_I f(\mathbf R_I)   \nonumber  \\
&= \mathrm e^{\mathrm i Z_I e f(\mathbf R_I)/\hbar} \,\overline \Pi_I\, \mathrm e^{-\mathrm i Z_I ef(\mathbf R_I)/\hbar}
\end{align}
and the Hamiltonian becomes
\begin{align}
    \tilde H - \mathrm i \hbar \frac{\partial}{\partial t} &= \left(\prod_{I=1}^{N_\text{nuc}} \mathrm e^{\mathrm i Z_I e f(\mathbf R_I)/\hbar} \right) \times \nonumber \\ & \times \left(H - \mathrm i \hbar \frac{\partial}{\partial t}\right) \left( \prod_{I=1}^{N_\text{nuc}} \mathrm e^{-\mathrm i Z_I e f(\mathbf R_I)/\hbar}\right).
\end{align}
Again, the gauge transformation is compensated for by the unitary transformation of the nuclear wave function in Eq.~\eqref{nucwf_gauge_trans}
and  does not affect any observable quantities as long as our computational model can perform such transformations.

\subsection{Equations of Motion}

We derive the equations of motion by considering the time evolution of the expectation values of the position and momentum operators using Eq.~\eqref{nucTDSE}, and noting that
\begin{align}
    \frac{\mathrm{d}}{\mathrm{d} t} \bra{\psi}  R_{J\beta} \ket{\psi} &= \frac{\mathrm{i}}{\hbar} \bra{\psi} [ H, R_{J\beta}] \ket{\psi}, \label{eq:Rt} \\
      \frac{\mathrm{d}}{\mathrm{d} t} \bra{\psi} \overline{\Pi}_{J\beta} \ket{\psi} & = \frac{\mathrm{i}}{\hbar}  \bra{\psi} [ H, \overline \Pi_{J\beta}] \ket{\psi}.
      \label{eq:Pt}
\end{align}
To determine the commutators of $R_{J\beta}$ and $\overline \Pi_{J\beta}$ with the Hamiltonian, it is useful first to evaluate their commutators with the momentum operators:
\begin{align}
 [\overline \Pi_{I\alpha}, R_{J\beta}] &= -\mathrm i \hbar \delta_{IJ}\delta_{\alpha\beta}, \\
 [\overline \Pi_{I\alpha}, \overline \Pi_{J\beta}] &= \mathrm i \hbar e Z_J \delta_{IJ} \epsilon_{\alpha\beta\gamma}B_\gamma + \mathrm i \hbar \Omega_{I\alpha J\beta},
 \label{eq:PP}
\end{align}
where $\epsilon_{\alpha\beta\gamma}$ is the Levi-Civita tensor and summation over the repeated index $\gamma$ is implied. We have also introduced the Berry curvature $\boldsymbol{\Omega}$ with elements
\begin{align}
    \Omega_{I\alpha J\beta} &= \nabla_{J\beta}\chi_{I\alpha} - \nabla_{I\alpha} \chi_{J\beta} \nonumber \\&= \mathrm i \hbar 
    \big[
    \braket{\nabla_{I\alpha} \phi}{\nabla_{J\beta} \phi} -  \braket{\nabla_{J\beta} \phi}{\nabla_{I\alpha} \phi}
    \big] .
    \label{eom_t006}
\end{align}
In evaluating the commutator in Eq.\,\eqref{eq:PP}, it is useful to express the magnetic vector potential as $A_\alpha(\mathbf R_I) = \frac{1}{2} \epsilon_{\alpha\beta\gamma} B_\beta R_{I\gamma}$ and
the geometric vector potential in the symmetric form
\begin{equation}
    \chi_{I\alpha} = \frac{\mathrm i \hbar}{2} \big[ \braket{\nabla_{I\alpha} \phi}{\phi} - \braket{\phi}{\nabla_{I\alpha} \phi} \big] \, ,
\end{equation}
which is valid since the geometric vector potential is real valued.  We note from Eq.\,\eqref{eom_t006} that the Berry curvature is an antisymmetric $3N_\text{nuc} \times 3N_\text{nuc}$ matrix
\begin{equation}
    \boldsymbol{\Omega}^\mathrm{T} = - \boldsymbol{\Omega}
    \label{omegaA}
\end{equation}
For each pair of atoms $I$ and $J$, $\boldsymbol \Omega_{IJ}$ is the $3\times 3$ matrix whose elements are given by $[\boldsymbol \Omega_{IJ}]_{\alpha\beta}= \Omega_{I\alpha J\beta}$ and we note the symmetry 
\begin{equation}
\boldsymbol \Omega_{IJ}^\mathrm T = - \boldsymbol \Omega_{JI}\,.
\label{omegaAIJ}
\end{equation}
Evaluating the gauge-transformed Berry curvature,
we obtain
\begin{align}
    \Omega_{I\alpha J\beta}^\prime &=  \nabla_{J\beta}\chi_{I\alpha}^\prime - \nabla_{I\alpha} \chi_{J\beta}^\prime \nonumber \\
    &=  \nabla_{J\beta}\left(\chi_{I\alpha} - \nabla_{I\alpha} F \right) - \nabla_{I\alpha}\left( \chi_{J\beta} - \nabla_{J\beta} F\right) \nonumber \\
    &= \nabla_{J\beta}\chi_{I\alpha} - \nabla_{I\alpha} \chi_{J\beta} = \Omega_{I\alpha J\beta},
\end{align}
demonstrating its gauge invariance.

Returning to Eqs.~\eqref{eq:Rt} and~\eqref{eq:Pt} and making
use of the identity $[a^2,b] = \{a, [a,b]\} = a[a,b] + [a,b]a$ where curly brackets denote the anti-commutator, we obtain the time derivatives
\begin{align}
    \frac{\mathrm{d}}{\mathrm{d} t} \bra{\psi}  R_{J\beta} \ket{\psi} &= \frac{1}{M_J} \bra{\psi} \overline{\Pi}_{J\beta} \ket{\psi},
    \\
      \frac{\mathrm{d}}{\mathrm{d} t} \bra{\psi} \overline{\Pi}_{J\beta} \ket{\psi} &= -\frac{\partial}{\partial R_{J\beta}} \bra{\psi}  U_\text{BO} \ket{\psi} 
        \nonumber \\
        &\phantom{=} -\frac{Z_J}{M_J} \epsilon_{\alpha\beta\gamma} \bra{\psi} \overline{\Pi}_{J\alpha} \ket{\psi} B_\gamma
        \nonumber \\
      &\phantom{=} - \sum_I \frac{1}{2M_I} \bra{\psi} \left\{ \overline \Pi_{I\alpha}, \Omega_{I\alpha J\beta} \right\} \ket{\psi}.
\end{align}
Finally, we assume that the nuclear wave function is sharply peaked around $\overline {\mathbf R}=\bra{\thewfnuc}\mathbf{R}\ket{\thewfnuc}$ so that
\begin{align}
\bra{\thewfnuc}\theebo\ket{\thewfnuc} &\approx U_\text{BO}(\overline {\mathbf R}),
\label{eom_t012} 
\\
\bra{\thewfnuc}\Omega_{I\alpha J\beta}\ket{\thewfnuc}& \approx \Omega_{I\alpha,J\beta}(\overline{\mathbf{R}}),
\label{eom_t014} 
\\
\bra{\thewfnuc} \{\overline \Pi_{I\alpha}, \Omega_{I\alpha J\beta}\} \ket{\psi} & \approx 2 \, \Omega_{I\alpha,J\beta}(\overline{\mathbf{R}}) \, \bra{\thewfnuc} \overline{\Pi}_{I\alpha} \ket{\psi}.
\label{eom_t016}
\end{align}
Alternatively, the last  approximation can also be obtained from a more restrictive combination of locality assumptions and a resolution of the identity, neglecting all couplings to excited states.

Using the locality assumptions in Eqs.~\eqref{eom_t012}--\eqref{eom_t016}, the equations of motion become
\begin{align}
M_{I} \, \ddot{ \overline {\mathbf R}}_{I}
= \mathbf F^{\text{BO}}_{I}(\overline{\mathbf{R}})
+\mathbf F^{\text{L}}_{I}(\overline{\mathbf{R}})
+\mathbf F^{\text{B}}_{I}(\overline{\mathbf{R}})
\label{eom_t015}
\end{align}
where we have introduced the Born--Oppenheimer force
\begin{align}
\mathbf{F}^{\text{BO}}_{I}(\overline {\mathbf{R}})
= - \boldsymbol \nabla_I  U_\text{BO}(\overline {\mathbf R}) 
\, ,
\label{BOforce}
\end{align}
the (bare) Lorentz force
\begin{equation}
\mathbf{F}^{\text{L}}_{I}(\overline {\mathbf{R}}) =- e Z_{I}  \, \mathbf B \times
\dot{ \overline {\mathbf{R}}}_{I} \, ,
\end{equation}
and the Berry (screening) force
\begin{equation}
\mathbf{F}^{\text{B}}_{I}(\overline {\mathbf{R}}) =\sum_{J} \boldsymbol{\Omega}_{IJ}( {\overline{\mathbf{R}}}) \, \dot{\overline{\mathbf{R}}}_{J} \, .
\end{equation}
where $\boldsymbol \nabla_I$ in~Eq.\,\eqref{BOforce} differentiates with respect 
to $\overline { \mathbf R}_{I}$. The screened Lorentz force on nucleus $I$ is the sum of the bare Lorentz force and the Berry force on this nucleus. Henceforth, we  omit the argument $\overline {\mathbf R}$ to the forces.

Equation\,\eqref{eom_t015} is legitimate under the assumptions stated, but will be inadequate in the presence of, for example, singularities such as conical intersections. There has been recent work seeking to address this issue in the context of dynamics equations where the Berry curvature is present.\cite{Rawlinson2020} Methods such as surface hopping\cite{Tully1990,Subotnik2016} can also be applicable in this framework. However, these concerns are beyond the scope of the present work

\subsection{Screened Lorentz Force}
\label{sec:BerryForce}

The screened Lorentz force on nucleus $I$ is calculated from the Berry curvature tensor, with a contribution from each atom in the molecule:
\begin{align}
\mathbf F^{\text{LB}}_{I} =
- e Z_{I} \mathbf{B} \times \dot{ \overline {\mathbf R}}_{I} +
\sum_{J} \boldsymbol \Omega_{IJ} \dot{\overline{\mathbf{R}}}_{J} 
\end{align}
We recall from Eq.\,\eqref{omegaA} that $\boldsymbol \Omega$ is anti-symmetric, implying that each three-by-three diagonal block $\boldsymbol \Omega_{II}$ is anti-symmetric, while  the off-diagonal three-by-three blocks $\boldsymbol \Omega_{IJ}$, with $I \neq J$, are in general not anti-symmetric. We note that the product of a three-by-three matrix with a vector may be expressed as a cross product if and only if the matrix is anti-symmetric
\begin{equation}
\begin{pmatrix} 
0 & - a_3 & a_2 \\ a_3 & 0 & -a_1 \\ -a_2 & a_1 & 0 
\end{pmatrix} \begin{pmatrix} b_1 \\ b_2 \\ b_3 \end{pmatrix} = 
\begin{pmatrix} a_1 \\ a_2 \\ a_3 \end{pmatrix} \times \begin{pmatrix} b_1 \\ b_2 \\ b_3 \end{pmatrix}
=\mathbf a \times \mathbf b,
\end{equation}
Therefore, decomposing the blocks of the Berry curvature tensor into symmetric and anti-symmetric parts,
\begin{equation}
\boldsymbol \Omega_{IJ}^\text{S} = \frac{1}{2} \left( \boldsymbol \Omega_{IJ} + \boldsymbol \Omega_{IJ}^\text T \right), \;
\boldsymbol \Omega_{IJ}^\text{A} = \frac{1}{2} \left( \boldsymbol \Omega_{IJ} - \boldsymbol \Omega_{IJ}^\text T \right),
\end{equation}
and introducing a vector consisting of the independent elements of the anti-symmetric part of the Berry curvature tensor 
\begin{equation}
\boldsymbol \omega_{IJ}^\text A = \begin{pmatrix} \Omega_{IzJy}^\text A \\ \Omega_{IxJz}^\text A \\ \Omega_{IyJx}^\text A \end{pmatrix},
\end{equation}
we may write the screened Lorentz force in the manner
\begin{align}
\mathbf{F}^\text{LB}_{I} =
- e Z_{I} \mathbf{B} \times \dot{\overline{\mathbf{R}}}_{I} & +
\sum_{J} \boldsymbol{\omega}_{IJ}^{\text{A}} \times \dot{ \overline {\mathbf{R}}}_{J} \nonumber \\
& + \sum_{J \neq I} \boldsymbol{\Omega}_{IJ}^\text{S} \dot{ \overline{\mathbf{R}}}_{J}
\label{eq:BerryForce}
\end{align}
where $\boldsymbol \Omega_{JJ}^\text S = \mathbf 0$. 

Taking the field vector $\mathbf B$ to define the $z$-axis and introducing the effective charges $Q_{IJ}$ by the condition
\begin{equation}
e\,Q_{IJ}  \mathbf B = -\boldsymbol \omega^\text{A}_{IJ}
\label{Qij}
\end{equation}
we obtain
\begin{align}
\mathbf{F}^\text{LB}_{I} =
-\sum_{J} e \left(\delta_{IJ} Z_I + Q_{IJ} \right) \mathbf{B} \times \dot{ \overline {\mathbf{R}}}_{J} 
+\sum_{J \neq I} \boldsymbol{\Omega}_{IJ}^\text{S} \dot{ \overline {\mathbf{R}}}_{J}
\label{eq:BerryForce1}
\end{align}
for the screened Lorentz force on nucleus $I$. For an atom, we have (in the limit of a complete one-electron basis) $Q_{11} = - N_\text{el}$, giving 
\begin{equation}
\mathbf F^\text{LB}_1 = e (Z_1 - N_\text{el}) \dot{ \overline {\mathbf R}}_1 \times \mathbf B
\quad \mbox{(atom)}
\label{e_atom_FLB}
\end{equation}
 where $Z_1 - N_\text{el}$ is the total charge of the atom. As we shall see in the next section, $Q_{11} = - N_\text{el}$ is satisfied even for an incomplete basis of London atomic orbitals. 
 
 \subsection{Berry curvature of a London atomic orbital}
 
Let $\varphi_{lm}(\mathbf r;\mathbf R)$ be a normalized real-valued field-free atomic orbital at position $\mathbf R$ of solid-harmonic quantum numbers $\ell$ and $m$ and let
 \begin{equation}
 \psi_{\ell m} (\mathbf r;\mathbf R) = \mathrm e^{-\mathrm i e \mathbf A(\mathbf R_1)\cdot \mathbf r /\hbar } \varphi_{\ell m}(\mathbf r;\mathbf R)
 \end{equation}
 be the corresponding field-dependent London atomic orbital, where $\mathbf A(\mathbf R) = \frac{1}{2} \mathbf B \times \mathbf R$ is the
 magnetic vector potential at the position of the orbital $\mathbf R$. The geometric vector potential generated by the London orbital is
 \begin{equation}
 \boldsymbol \chi_{\ell m}(\mathbf R) = \int\!\! \psi_{\ell m}^\ast \mathbf P \psi_{\ell m}\, \mathrm d \mathbf r = e \mathbf A(\mathbf R)\,,
 \label{eq:chiint}
 \end{equation}
 where $\mathbf P = - \mathrm i \hbar \boldsymbol \nabla_\mathbf R$.
To see how this result arises, we first note that (by rearranging the scalar triple product)
 \begin{equation}
 \boldsymbol \nabla_\mathbf R \left( - \mathbf A(\mathbf R)\cdot \mathbf r \right) = 
  \boldsymbol \nabla_\mathbf R \left(  \mathbf A(\mathbf r)\cdot \mathbf R \right) = 
 \mathbf A(\mathbf r). 
 \end{equation}
The integrand of Eq.\,\eqref{eq:chiint} may therefore be written as
 \begin{align}
 \psi_{\ell m}^\ast \mathbf P \psi_{\ell m} &=  e \mathbf A(\mathbf r) \varphi_{\ell m}^\ast \varphi_{\ell m} +
\varphi_{\ell m}^\ast  \mathbf P \varphi_{\ell m} \nonumber \\
&=  e \mathbf A(\mathbf R) \varphi_{\ell m}^2 + e \mathbf A(\mathbf r_\mathbf R) \varphi_{\ell m}^2 + \tfrac{1}{2}
 \mathbf P \varphi_{\ell m}^2
 \end{align}
where, in the last step, we have introduced $\mathbf r_\mathbf R = \mathbf r - \mathbf R$ and used the fact that $\varphi_{\ell m}$ is real valued. Integrating over all space, the first term gives $e \mathbf A(\mathbf R)$ by normalization of the orbital, the second term vanishes by parity symmetry, while the last term vanishes by the normalization of the atomic orbital (having interchanged differentiation and integration), yielding Eq.~\,\eqref{eq:chiint}.

From the definition of the Berry curvature tensor given in Eq.~\eqref{eom_t006}, we obtain
\begin{equation}
\boldsymbol \Omega_{\ell m} = e \begin{pmatrix} 0 & -B_z & B_y \\ B_z & 0 & - B_x \\ -B_y & B_x & 0  
\end{pmatrix}
\end{equation}
and $\boldsymbol \omega_{\ell m}^\text A = e \mathbf B$. Assuming a one-electron system described by this London orbital, the effective charge is then $-e$. For antisymmetric product of $N_\text{el}$ London spin orbitals, the effective charge is $- N_\text{el} \, e$.

\subsection{Berry Curvature from Finite Difference}
The elements of the Berry curvature are combinations of the overlap of wave-function derivatives; see Eq.\,\eqref{eom_t006}. Our task is to calculate these derivative overlaps from
finite differences:
\begin{align}
& \braket{\nabla_{I\alpha}\phi}{\nabla_{J\beta} \phi} \approx
\frac{S^{++}_{I\alpha J\beta}\!-\!S^{+-}_{I\alpha J\beta}\!-\! S^{-+}_{I\alpha J\beta} \!+\! S^{--}_{I\alpha J\beta}}{4\delta_{I\alpha}\delta_{J\beta}} \,.
\label{fd_000}
\end{align}
We have here introduced the notation
\begin{align}
S^{\pm \pm}_{I\alpha J\beta} = \braket{\phi_{\pm I\alpha}}{\phi_{\pm J\beta}} 
\label{fd_000a}
\end{align}
where
\begin{align}
\ket{\thewfel_{\pm I\alpha}} = \ket{\thewfel(\mathbf{r}; \mathbf R \pm \boldsymbol \delta_{I\alpha})}
\label{fd_001}
\end{align}
is the electronic wave function at the geometry $\mathbf R \pm \boldsymbol \delta_{I\alpha}$ where
$\boldsymbol \delta_{I\alpha}$ denotes a perturbation of magnitude $\delta_{I\alpha}$ in the $I\alpha$ component of $\mathbf R$.

The finite-difference method has been used previously to calculate contributions to the energy from the diagonal Born--Oppenheimer correction for real-valued wave functions.\cite{Cencek1997,Valeev2003,Schneider2019} The present work follows the same finite-difference formulation for the calculation of the Berry curvature, but with two important complicating factors. The first is that the Berry curvature tensor contains non-diagonal derivative overlaps with respect to the nuclear coordinates, which leads to the absence of simplifications that would otherwise be present in the diagonal case (e.g., the form of the numerator in Eq.\,\eqref{fd_000}). The second and more difficult complication is that, in the presence of a magnetic field, the electronic wave function is in general complex. For real-valued wave functions, the global phase angle $\lambda$ is constrained to be $\lambda_n=n\pi$ where $n$ is an integer, which means that any perturbed wave function
\begin{align}
\mathrm e^{- \mathrm i \lambda_n/\hbar}\ket{\thewfel} \rightarrow \mathrm e^{- \mathrm i \lambda_m/\hbar}\ket{\thewfel_{\pm I\alpha}}
\label{fd_002}
\end{align}
may at most undergo a sign change as a result of the change in phase. In the complex case, the phase angle becomes a continuous variable and each separate perturbation may result in a drastically different and uncontrolled phase. For example, the derivative of the electronic wave function with respect to a nuclear coordinate can be represented via finite difference according to
\begin{align}
   \ket{\nabla_{I\alpha} \phi}  &\approx \frac{\ket{\tilde \phi_{+I\alpha}} - \ket{\tilde \phi_{-I\alpha}}}{2\delta_{I\alpha}}
      \nonumber \\
   &= \frac{\ket{\phi_{+I\alpha}}\mathrm e^{-\mathrm i\lambda_{+}} - \ket{\phi_{-I\alpha}} \mathrm e^{-\mathrm i\lambda_{-}}}{2\delta_{I\alpha}}
\label{erik_0}
\end{align}
where tilde denotes a phase-corrected wave function.

We want to evaluate Eq.~\eqref{erik_0} with the phase-corrected wave function \smash{$\tilde \phi_{\pm I\alpha}$} since the raw, uncorrected wave function $\phi_{\pm I\alpha}$ gives an ill-defined gradient. For some values of $\lambda_{\pm}$, the numerator in Eq.\,\eqref{erik_0} is on the order of $\delta_{I\alpha}$ but for most it is on the order of one. Moreover, since $\phi_{\pm I\alpha}$ has an uncontrolled phase, we cannot simply set $\lambda_{\pm} = 0$. It is therefore necessary to perform the finite difference scheme accounting for the arbitrary phase of each perturbed wave function in a justified, systematic manner.

Towards this end, we note that the overlap of any perturbed wave function with the reference wave function can be represented in polar form by
\begin{equation}
\braket{\thewfel}{\thewfel_{\pm I\alpha}} = \eta_{\pm I\alpha} \mathrm e^{\mathrm i\lambda_{\pm I\alpha}},
\label{fd_003}
\end{equation}
where $\eta_{\pm I\alpha}$ is the modulus and $\lambda_{\pm I\alpha}$ the argument. 
We now multiply $\ket{\thewfel_{\pm I\alpha}}$ by a phase factor that renders the result of Eq.\,\eqref{fd_003} real. This is accomplished according to   
\begin{equation}
\ket{\thewfel_{\pm I\alpha}} \rightarrow \mathrm e^{-\mathrm i\lambda_{\pm I\alpha}} \ket{\thewfel_{\pm I\alpha}}
= \vert \tilde \phi_{\pm I \alpha} \rangle,
\label{fd_004} 
\end{equation}
which gives
\begin{equation}
\braket{\thewfel}{\tilde{\phi}_{\pm I\alpha}} = \eta_{\pm I\alpha}.
\label{fd_005}
\end{equation}
The condition on the perturbed wave functions imposed by Eq.~\eqref{fd_005} is (in the context of finite difference) tantamount to a gauge transform that causes the geometric vector potential to vanish. To see this, we write the finite difference approximation to the geometric vector potential in the form
\begin{align}
\tilde \chi_{I\alpha} &\approx \dfrac{\mathrm i\hbar\left(\braket{\thewfel}{\tilde \phi_{-I\alpha}} - 
\braket{\thewfel}{\tilde \phi_{+I\alpha}}\right)}{2\delta_{I\alpha}} \nonumber \\
&= \dfrac{\mathrm i\hbar\left(\eta_{-I\alpha} - \eta_{+I\alpha}\right)}{2\delta_{I\alpha}} = 0,
 \label{fd_006} 
\end{align}
as follows from the fact that $\chi_{I\alpha}$ is real-valued and since Eq.\,\eqref{fd_006} is pure imaginary, it must be zero. As shown in Section C, the Berry curvature is invariant under geometric gauge transformations, and is thus unaffected. We are now in a position to calculate the elements of the Berry curvature tensor using the phase corrections given in Eq.\,\eqref{fd_004} so that
\begin{align}
&\left\langle \nabla_{I\alpha} \phi \left\vert \nabla_{J\beta}\phi  \right.\right\rangle \approx
\frac{\tilde S^{++}_{I\alpha J\beta}\! -\! \tilde S^{+-}_{I\alpha J\beta}\! - \!\tilde S^{-+}_{I\alpha J\beta} \!+\! \tilde S^{--}_{I\alpha J\beta}}{4\delta_{I\alpha}\delta_{J\beta}},
\label{fd_000ay}
\end{align}
with
\begin{align}
\tilde S^{\pm \pm}_{I\alpha J\beta} &= \langle \tilde \phi_{\pm I\alpha} \vert \tilde \phi_{\pm J \beta}\rangle \nonumber \\
& = \mathrm e^{\mathrm i \lambda_{\pm I \alpha}} \langle  \phi_{\pm I\alpha} \vert  \phi_{\pm J \beta}\rangle \mathrm e^{-\mathrm i \lambda_{\pm J \beta}},
\label{fd_000ax} 
\end{align}
where the phase factors are given in Eq.~\eqref{fd_003}. 

\section{Results}

In this section, we present calculations of the Berry curvature for the hydrogen, helium, and lithium atoms, as well as the H$_2$ and CH$^+$ molecules using the finite-difference scheme outline in Section II. All calculations in this work were performed with the software package {\sc london}.\cite{LondonProgram} London orbitals were used in all calculations in Section III.B and Section III.C, ensuring gauge-origin invariance. Calculations in Section III.A were performed without using London orbitals.  

\subsection{H, He, and Li Atoms}

In order to demonstrate the importance of London orbitals, the Berry curvature was calculated for the hydrogen, helium, and lithium atoms as a function of basis set size without London orbitals, for fields strengths of $|\theb| = 0.1B_0$ and $|\theb| = 1.0B_0$, where $B_0 = 2.35 \times 10^5$\,T is one atomic unit magnetic field strength. Calculations were performed with Dunning's correlation-consistent polarized valence basis sets cc-pV$X$Z\cite{ccpVXZ} (without diffuse functions) and aug-cc-pV$X$Z\cite{augccpVXZ} (with diffuse functions), in both cases with cardinal numbers $2 \leq X \leq 4$. In addition, we performed calculations in the minimal STO-3G basis\cite{sto3g}, with cardinal number $X=1$.

\begin{figure*}[h]
\centering
\begin{tabular}{ll}
(a) & (b) \\
\includegraphics[width=0.48\textwidth]{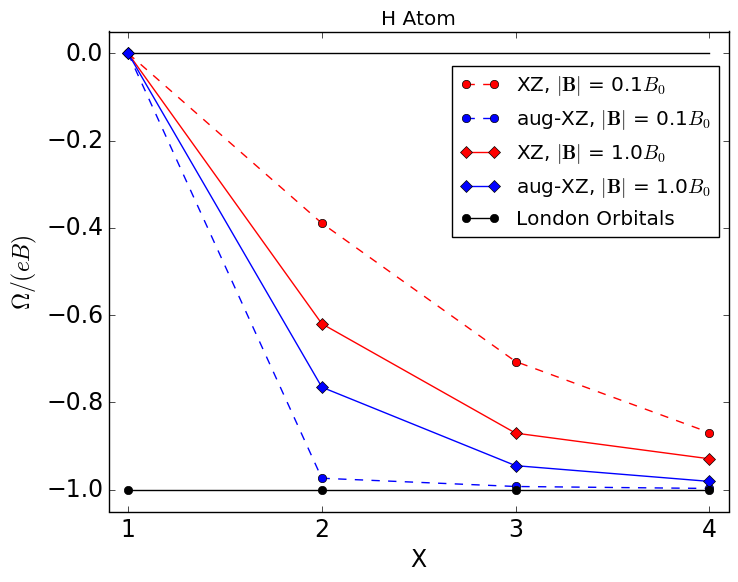} &
\includegraphics[width=0.48\textwidth]{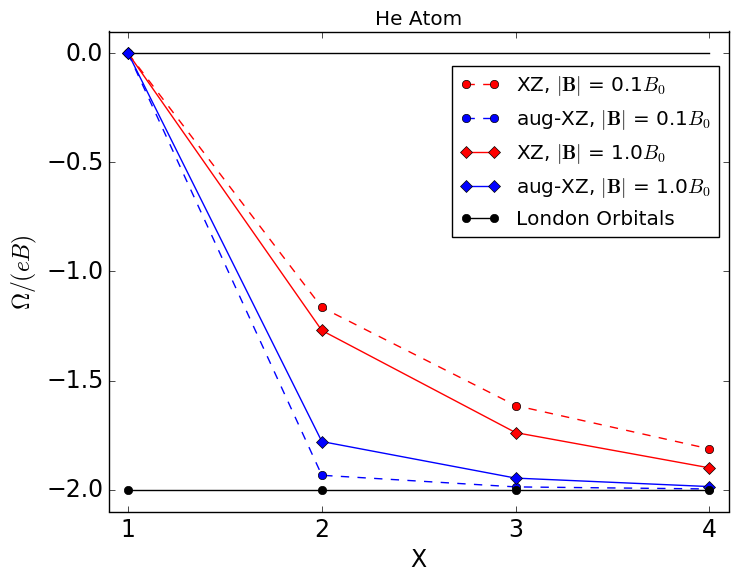} \\
(c) \\
\includegraphics[width=0.48\textwidth]{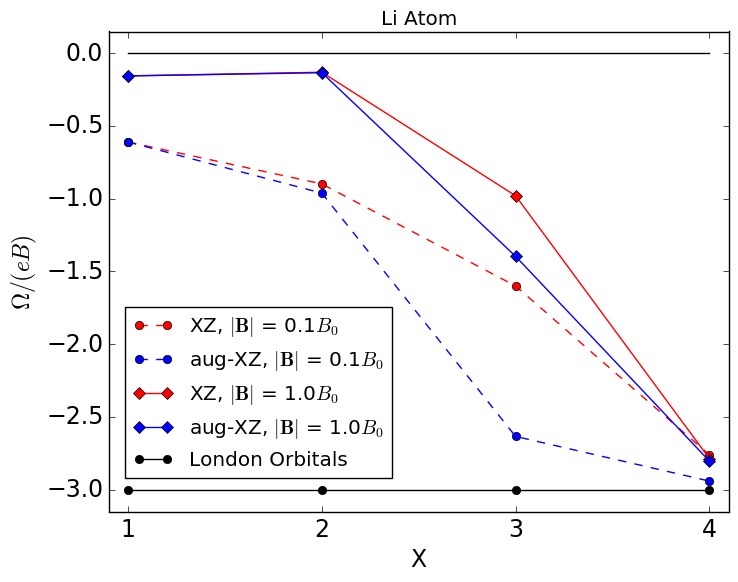}
\end{tabular}
\caption{Electronic charges calculated from the Berry curvature as a function of basis set without London orbitals for the hydrogen (a), helium (b) and lithium (c) atoms. Lithium values were calculated using UHF with a spin-projection corresponding to a doublet. Data is plotted as a function of basis set cardinal number $X$ for the cc-pVXZ and aug-cc-pVXZ basis sets, as well as the STO-3G basis set ($X=1$). The magnetic field was oriented along the z-axis, with a strength of $|\theb| = 0.1B_0$ and $|\theb| = 1.0B_0$  as reflected in the plots. London orbitals reproduce the exact electronic charge for all basis sets/magnetic field strengths.}
\label{fig_atoms_cbs}
\end{figure*}

For neutral atoms such as those considered here, we recall from Eq.\,\eqref{e_atom_FLB} that the exact screened Lorentz force is zero because of the complete cancellation of nuclear and electronic charges in neutral systems. Importantly, \emph{this cancellation is achieved exactly when calculating the Berry curvature with London orbitals, in an orbital basis of any size}. By contrast, in a finite basis without  London phase factors, only partial cancellation is achieved---see Figure\,\ref{fig_atoms_cbs}, where we have plotted the electronic charges of H, He, and Li calculated from the Berry curvature.

As seen in these plots, convergence of electronic charge calculated from the Berry curvature without using London orbitals is slow and irregular, although slightly faster when diffuse functions are added to the basis sets. In fact, a quantitative agreement with the number of electrons in the atom is only achieved with the aug-cc-pVQZ basis set and then only for H and He. Because of the presence of core electrons (for which the basis sets used here have only single-zeta quality), the basis-set convergence for Li is slower than for H and He. In general, we expect the basis-set convergence of the Berry curvature in atomic calculations without London orbitals to become more difficult with an increasing number of electrons. In molecules, where in addition the atoms may be far away from the gauge origin, convergence will be slower still.

For the calculation of Berry curvature, it is clearly essential to use London orbitals, which give the correct number of electrons even in the smallest basis sets.
In all molecular calculates presented here, we therefore use London orbitals.    

\subsection{H$_2$ Molecule}

The Berry curvature of the lowest singlet and $\beta\beta$ triplet states of H$_2$ has been calculated at the Hartree--Fock level of theory, at orientations of the molecule both parallel and perpendicular to the magnetic field. Calculations were performed with the STO-3G, cc-pVDZ, and cc-pVTZ basis sets of London atomic orbitals for magnetic field strengths of 0.1$B_0$ and 1.0$B_0$.

It should be noted that linear dependence in the basis set hampers investigations for small bond distance. For this reason, the Berry curvature was calculated down to a bond distance limit of $0.1$\,bohr, beyond which linear dependence starts to manifest itself in the cc-pVTZ basis. The finite-difference step size used in all calculations is $5.0 \times 10^{-4}$~bohr, in combination with a DIIS convergence tolerance of $1.0 \times 10^{-8}$ a.u. This step size has been shown to be robust and effective to a high level of accuracy in calculations of the DBOC,\cite{Cencek1997,Valeev2003} and we find that this is also the case for the Berry curvature, with step sizes between $1.0 \times 10^{-3}$ and $1.0 \times 10^{-4}$~bohr giving comparable results in most instances.

For H$_2$ with the a uniform magnetic field along the $z$-axis and molecular orientation either parallel or perpendicular to to the field, the Berry curvature takes the form
\begin{equation}
    \boldsymbol \Omega_{\text{H}_2} = \begin{pmatrix} \boldsymbol \Omega_{11} & \boldsymbol \Omega_{12} \\ \boldsymbol \Omega_{21} & \boldsymbol \Omega_{22} \end{pmatrix}=
    \left(\begin{array}{ccc|ccc}
    0 & -\kappa & 0 & 0 & - \tau & 0 \\ \kappa & 0 & 0 & \tau & 0 & 0 \\
    0&0&0&0&0&0\\\hline
    0& - \tau& 0 & 0& -\kappa & 0 \\
    \tau & 0 & 0 & \kappa & 0 & 0 \\
      0&0&0&0&0&0
    \end{array}\right)
    \label{res_000}
\end{equation}
where $\mathbf{\Omega}_{11} = \mathbf{\Omega}_{22}$ and $\mathbf{\Omega}_{12} =-\boldsymbol \Omega_{21}^\mathrm T= \mathbf{\Omega}_{21}$ 
and $\kappa$ and $\tau$ are obtained from Eq.\,\eqref{eom_t006}. Since
in this particular case all blocks are anti-symmetric, we may as discussed in Section\;\ref{sec:BerryForce} calculate the Berry force in cross-product form from 
\begin{equation}
\boldsymbol \omega_{11}^\text A = \boldsymbol \omega_{22}^\text A = \begin{pmatrix} 0 \\ 0 \\ \kappa \end{pmatrix}, \quad
\boldsymbol \omega_{12}^\text A = \boldsymbol \omega_{21}^\text A = \begin{pmatrix} 0 \\ 0 \\ \tau \end{pmatrix},
\end{equation}
yielding the equations of motion
\begin{align}
M_{I} \ddot{\overline{\mathbf{R}}}_{I}
&= \mathbf{F}^{\text{BO}}_{I} - 
\sum_{J} \left(e \delta_{IJ} Z_{I} \mathbf{B} - \boldsymbol{\omega}^{\text{A}}_{IJ} \right)
\times \dot{ \overline {\mathbf R}}_{J}.
\label{res_010}
\end{align}
Introducing screening charges $Q_{IJ}$ as given in Eq.\,\eqref{Qij}, we obtain
\begin{align}
Q_{II} = -\dfrac{\kappa}{e B_z}, \quad
Q_{IJ} = -\dfrac{\tau}{e B_z}, 
\label{res_012}
\end{align}
the equations of motion take the form
\begin{align}
M_{I} \ddot{\overline{\mathbf{R}}}_{I} 
& = \mathbf{F}^{\text{BO}}_{I} - 
\sum_{J} e \left(\delta_{IJ} \thezi + Q_{IJ}\right) \mathbf{B}
\times \dot{\overline{\mathbf{R}}}_{J},
\label{res_010x}
\end{align}
where the screening charges add up to the partial charge on each atom
\begin{align}
q_{I} = \sum_{J} Q_{IJ} = \sum_J Q_{JI}.
\label{res_013}
\end{align}
In general, however, the symmetric part of the Berry curvature does not vanish and we cannot express the force in terms of screening charges.

Our equations of motion for H$_2$ agree with those in Ref.\,\onlinecite{Ceresoli2007}, which are specialized to the case of vanishing symmetric component of the Berry curvature. The definition in Ref.\,\onlinecite{Ceresoli2007} of the
Berry curvature corresponds to
\begin{equation}
  \begin{split}
\Omega^{(IJ)}_\alpha & = -2 \hbar \, \mathrm{Im}\big( \epsilon_{\alpha\beta\gamma} \braket{ \nabla_{I\beta}\phi }{ \nabla_{J\gamma} \phi}\big)
        = \epsilon_{\alpha\beta\gamma} \Omega^A_{I\beta J\gamma},
  \end{split}
\label{res_003}
\end{equation}
and hence  $\boldsymbol{\Omega}^{(IJ)} = -2 \boldsymbol \omega_{IJ}^{\text{A}}$. Their
(anti-symmetric) Berry curvature thus differs from ours by a sign convention and a mistaken factor of two.

\begin{figure*}[h]
\centering
\begin{tabular}{ll}
(a) & (b) \\
\includegraphics[width=0.48\textwidth]{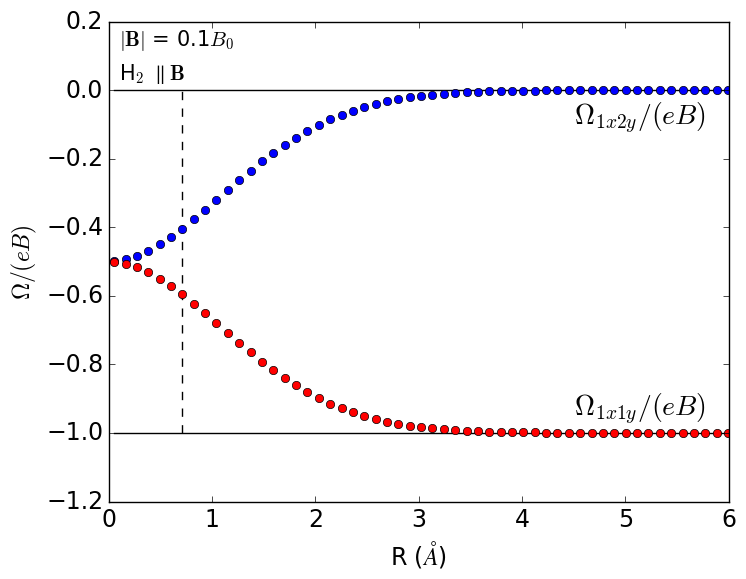} &
\includegraphics[width=0.48\textwidth]{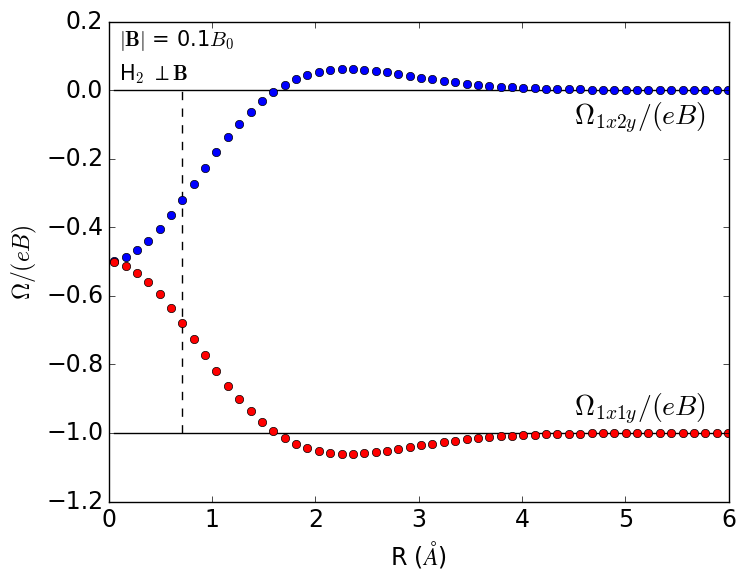} \\
(c) & (d) \\
\includegraphics[width=0.48\textwidth]{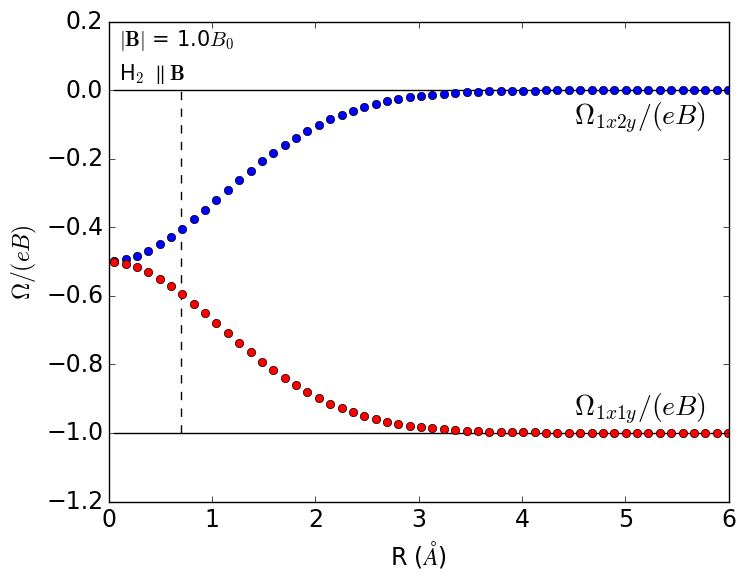} &
\includegraphics[width=0.48\textwidth]{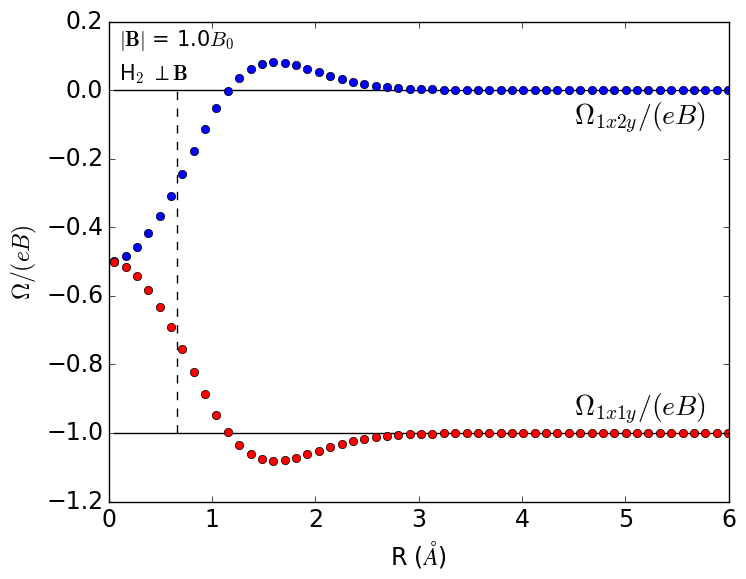} \\
\end{tabular}
\caption{Berry curvature of H$_2$ for the RHF singlet state with the STO-3G basis set for orientations parallel (a $\&$ c) and perpendicular (b $\&$ d) to the magnetic field. The field is oriented along the $z$-axis with a strength of $|\theb| = 0.1B_0$ (a $\&$ b) and $|\theb| = 1.0B_0$ (c $\&$ d). Molecular orientation perpendicular to the field is along the $x$-axis. Equilibrium bond distances are 0.712 $\text{\AA}$ (a), 0.711 $\text{\AA}$ (b), 0.698 $\text{\AA}$ (c), and 0.662 $\text{\AA}$ (d) as shown by the vertical dashed lines in each plot.}
\label{fig_sto3g-nobas-sto3g_0.1_1.0}
\end{figure*}

In Fig.\,\ref{fig_sto3g-nobas-sto3g_0.1_1.0}, we have for the
singlet ground state of H$_2$ at the RHF/STO-3G level of theory plotted the independent elements of the Berry curvature 
\begin{equation}
\Omega_{1x1y}=-\kappa, \quad \Omega_{1x2y} = -\tau,
\end{equation}
as a function of the bond distance for field strengths of 0.1$B_0$ and 1.0$B_0$, with the magnetic field oriented along the $z$-axis and the molecule oriented either parallel or perpendicular to the field.
These plots are in good agreement with the top panels in Fig.\,1 of Ref.\,\onlinecite{Ceresoli2007}, although a change in the curvature is noted moving from 0.1$B_0$ to 1.0$B_0$ as the stronger field begins to compress the molecule. For magnetic field strengths ranging from weak field to about 0.1$B_0$, the values of $\Omega/|\mathbf{B}|$ do not vary appreciably as a function of bond distance for the STO-3G RHF case. 

\begin{figure*}[h]
\centering
\begin{tabular}{ll}
(a) & (b) \\
\includegraphics[width=0.48\textwidth]{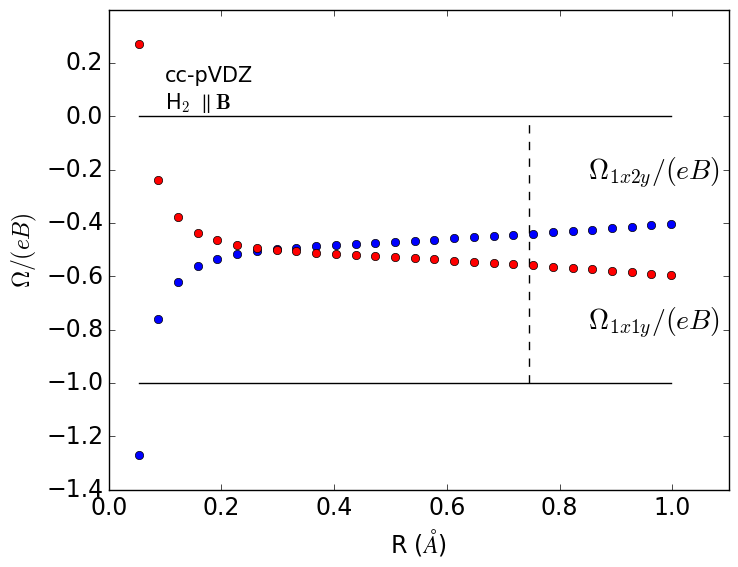} &
\includegraphics[width=0.48\textwidth]{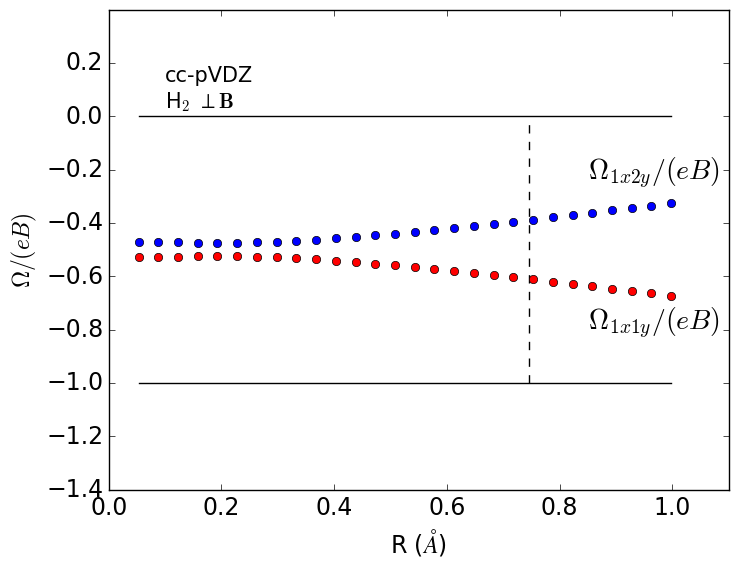} \\
(c) & (d) \\
\includegraphics[width=0.48\textwidth]{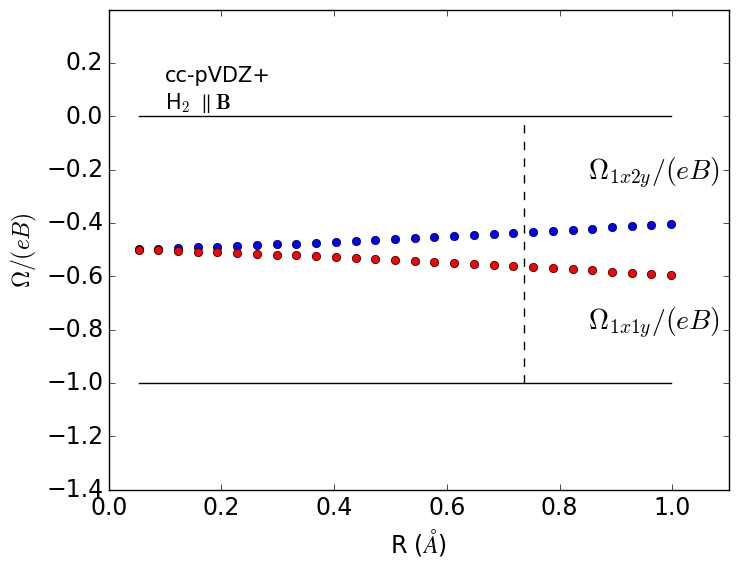} &
\includegraphics[width=0.48\textwidth]{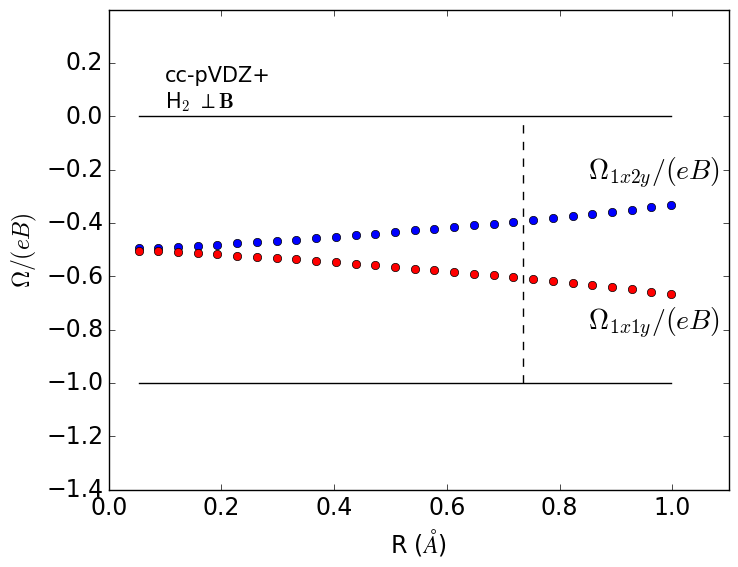} \\
\end{tabular}
\caption{Berry curvature of H$_2$ for the RHF singlet state with the cc-pVDZ basis set (a $\&$ b) and the hybrid cc-pVDZ+ basis set (c $\&$ d) for orientations parallel (a $\&$ c) and perpendicular (b $\&$ d) to the magnetic field. The field is oriented along the $z$-axis with a strength of $|\theb| = 0.1B_0$. for all panels. The cc-pVDZ+ basis is comprised of cc-pVDZ hydrogen basis functions centered at the nuclei and a cc-pVDZ Helium ghost center placed at the center of mass. Molecular orientation perpendicular to the field is along the $x$-axis. Equilibrium bond distances are 0.746 $\text{\AA}$ (a), 0.745 $\text{\AA}$ (b), 0.736 $\text{\AA}$ (c), and 0.735 $\text{\AA}$ (d) as shown by the vertical dashed lines in each plot.}
\label{fig_dz-dz-dz_0.1_close}
\end{figure*}

At very short bond distances, the calculated values of the Berry curvature are sensitive to the basis set. In particular, at bond distances less than 0.3~\AA, the convergence to the zero bond-distance limit is not the same across the STO-3G, cc-pVDZ, and cc-pVTZ basis sets. While the STO-3G basis set converges to a value of $-0.5eB_0$ for both $\Omega_{1x1y}$ and $\Omega_{1x2y}$, this behavior is not exhibited by the cc-pVDZ or cc-pVTZ basis sets, depending on the orientation to the field. Augmentation of the cc-pVDZ and cc-pVTZ hydrogen basis sets with a cc-pVDZ helium ghost basis at the center of mass greatly altered the convergence behavior in the region of small bond distance; see Fig.\,\ref{fig_dz-dz-dz_0.1_close} for the cc-pVDZ basis set. Such hybrid hydrogen/helium basis sets, denoted cc-pVDZ+ and cc-pVTZ+, have been used in subsequent Berry curvature calculations. The presence of the He ghost basis does not play a role at large internuclear separation and the dissociation limit is the same for basis sets with or without the ghost basis. 

\begin{figure*}[h]
\centering
\begin{tabular}{ll}
(a) & (b) \\
\includegraphics[width=0.48\textwidth]{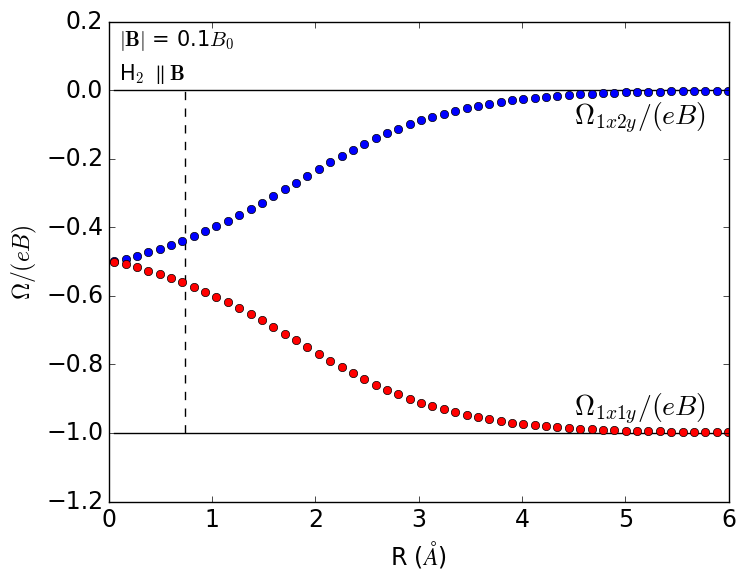} &
\includegraphics[width=0.48\textwidth]{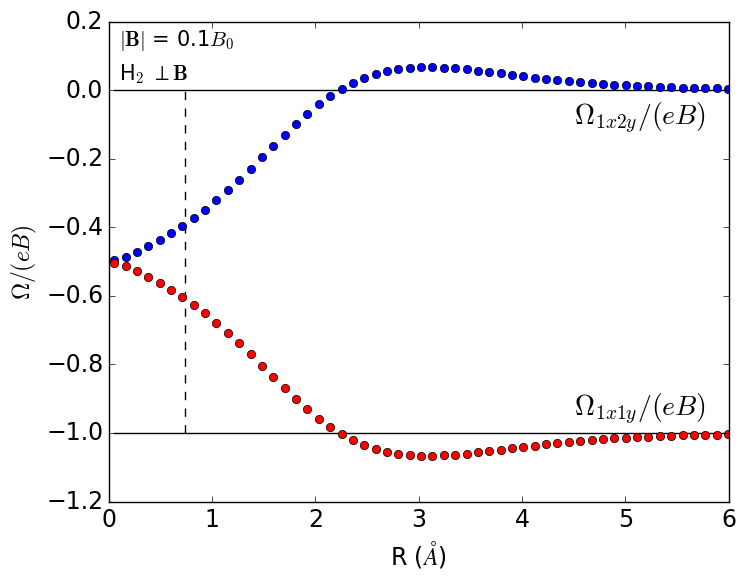} \\
(c) & (d) \\
\includegraphics[width=0.48\textwidth]{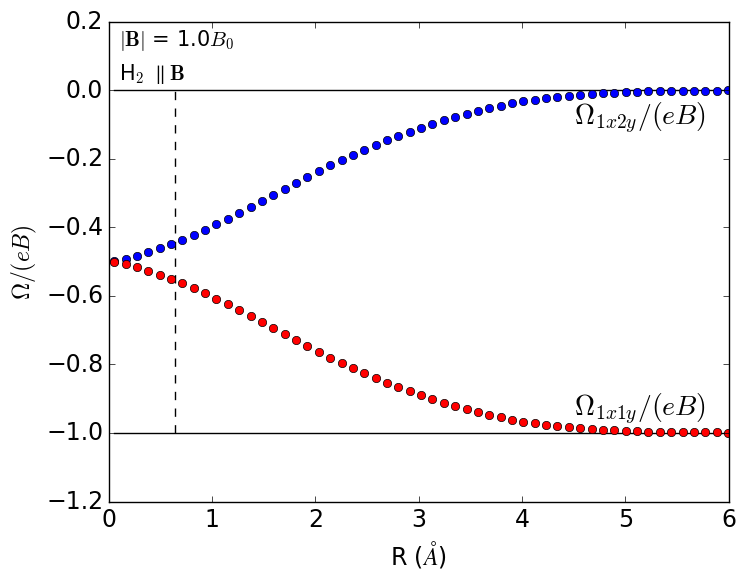} &
\includegraphics[width=0.48\textwidth]{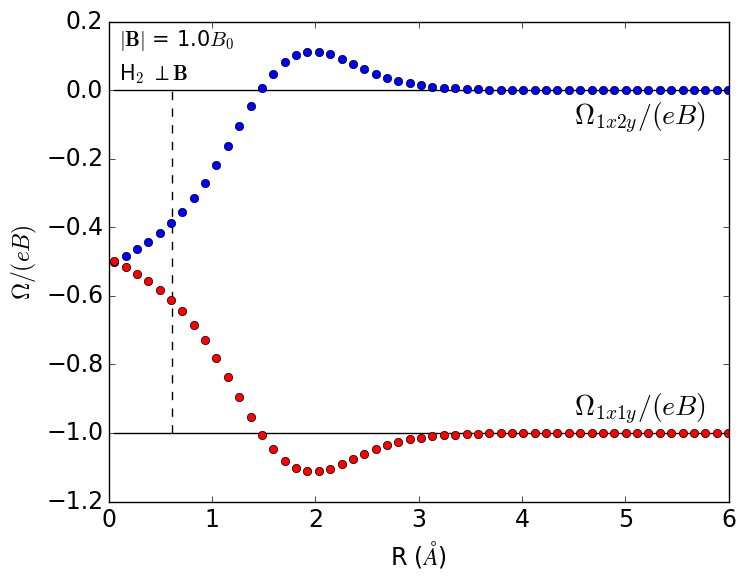} \\
\end{tabular}
\caption{Berry curvature of H$_2$ for the RHF singlet state with the cc-pVDZ+ basis set for orientations parallel (a $\&$ c) and perpendicular (b $\&$ d) to the magnetic field. The field is oriented along the $z$-axis with a strength of $|\theb| = 0.1B_0$ (a $\&$ b) and $|\theb| = 1.0B_0$ (c $\&$ d). The cc-pVDZ+ basis is comprised of cc-pVDZ hydrogen basis functions centered at the nuclei and a cc-pVDZ Helium ghost center placed at the center of mass. Molecular orientation perpendicular to the field is along the $x$-axis. Equilibrium bond distances are 0.736 $\text{\AA}$ (a), 0.735 $\text{\AA}$ (b), 0.645 $\text{\AA}$ (c), and 0.613 $\text{\AA}$ (d) as shown by the vertical dashed lines in each plot.}
\label{fig_dz-dz-dz_0.1_1.0}
\end{figure*}

In Fig.\,\ref{fig_dz-dz-dz_0.1_1.0}, we have plotted the RHF/cc-pVDZ+ Berry curvature against the bond distance for the ground-state H$_2$ molecule at field strengths 0.1$B_0$ and 1.0$B_0$. The  trend exhibited by the STO-3G basis is also observed for the cc-pVDZ+ basis, and  for cc-pVTZ+ as shown in the SI. Antiscreening and superscreening effects are observed for the perpendicular orientation of H$_2$ across all basis sets, where at certain bond distances the values of $\Omega_{1x2y}$ become positive (antiscreening) and the values of $\Omega_{1x1y}$ become less than $-1 eB_0$ (superscreening). The reproduction of this trend across all basis sets and magnetic fields strengths investigated here suggests it is not an artifact of the basis but a feature of the Berry curvature in this case.

\begin{figure*}[h]
\centering
\begin{tabular}{ll}
(a) & (b) \\
\includegraphics[width=0.48\textwidth]{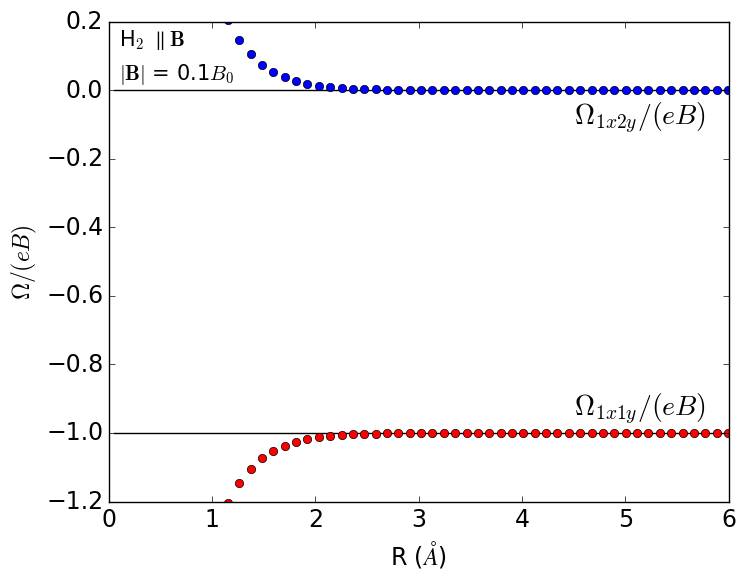} &
\includegraphics[width=0.48\textwidth]{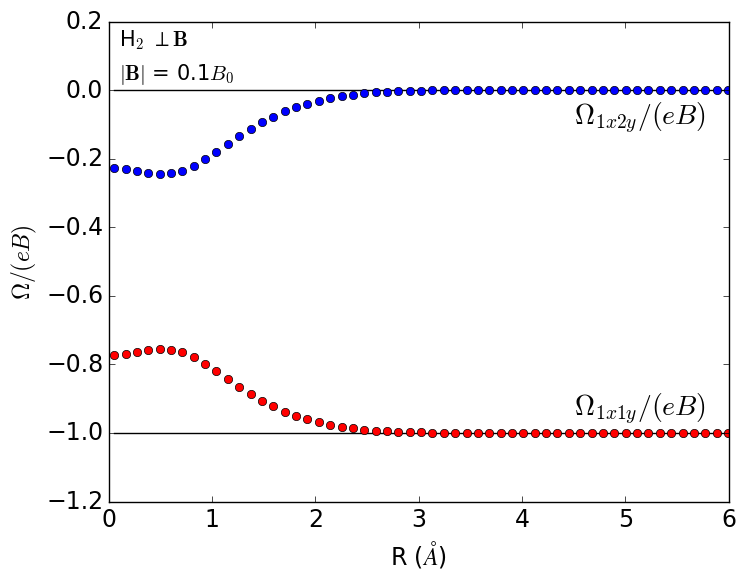} \\
(c) & (d) \\
\includegraphics[width=0.48\textwidth]{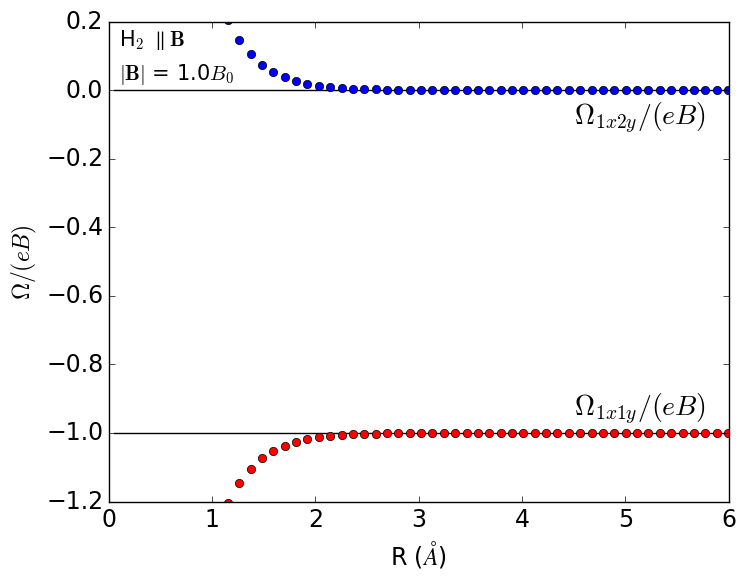} &
\includegraphics[width=0.48\textwidth]{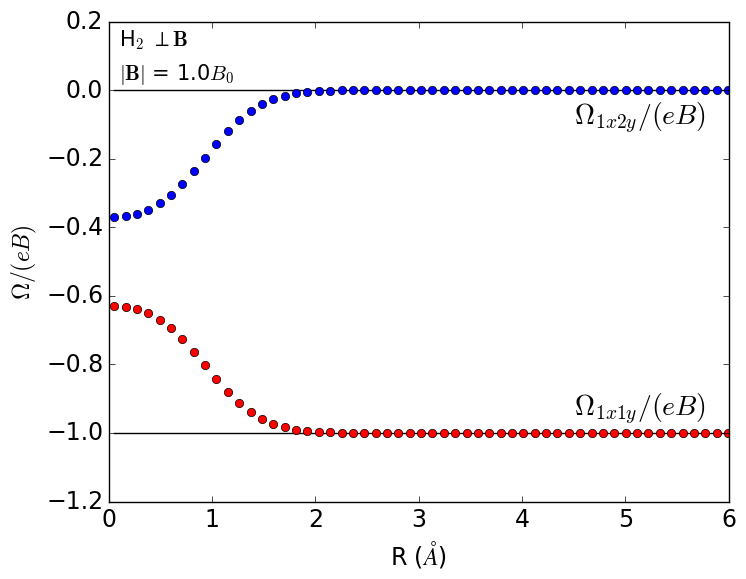} \\
\end{tabular}
\caption{Berry curvature of H$_2$ for the UHF $\beta\beta$ triplet state calculated with the STO-3G basis set for orientations parallel (a $\&$ c) and perpendicular (b $\&$ d) to the magnetic field. The field is oriented along the $z$-axis with a strength of $|\theb| = 0.1B_0$ (a $\&$ b) and $|\theb| = 1.0B_0$ (c $\&$ d). Molecular orientation perpendicular to the field is along the $x$-axis.}
\label{fig_uhf_sto3g-nobas-sto3g_0.1_1.0}
\end{figure*}

\begin{figure*}[h]
\centering
\begin{tabular}{ll}
(a) & (b) \\
\includegraphics[width=0.48\textwidth]{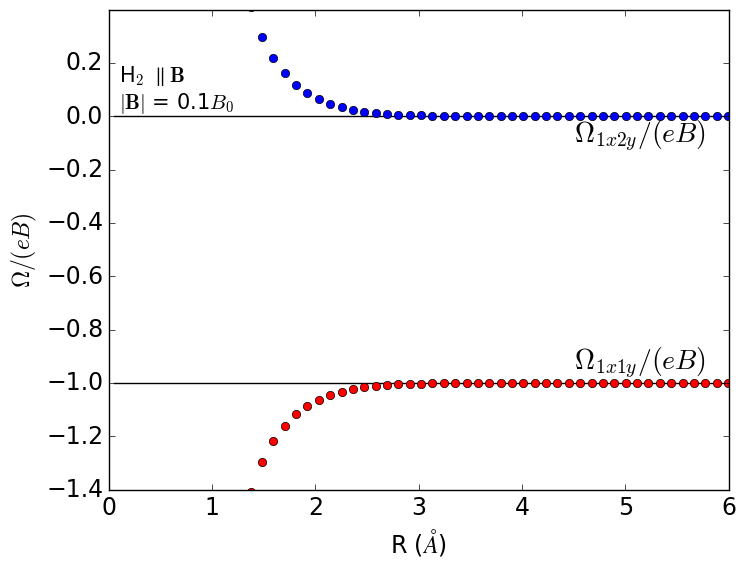} &
\includegraphics[width=0.48\textwidth]{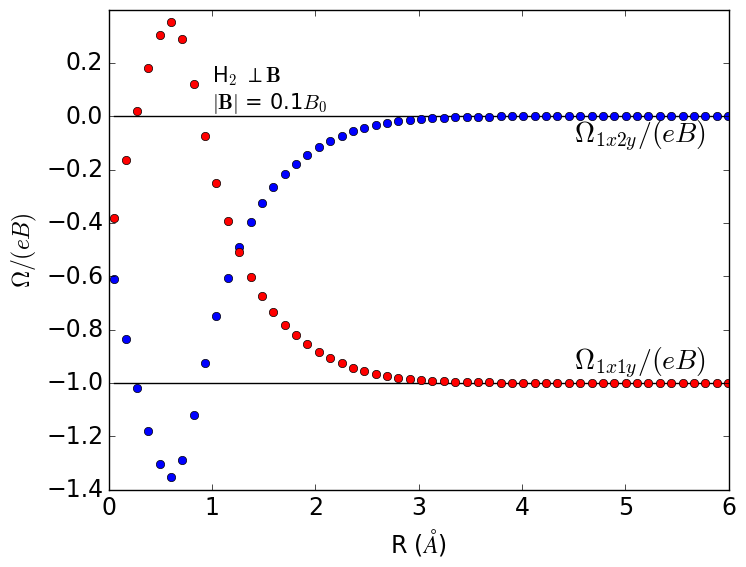} \\
(c) & (d) \\
\includegraphics[width=0.48\textwidth]{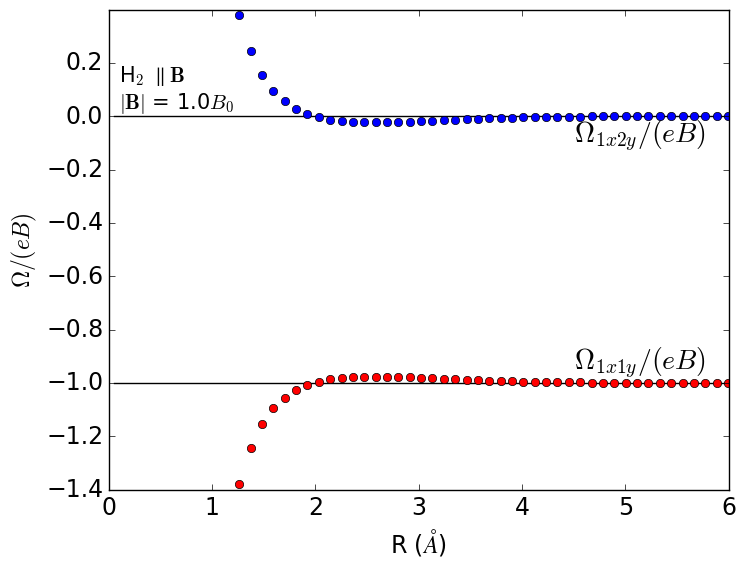} &
\includegraphics[width=0.48\textwidth]{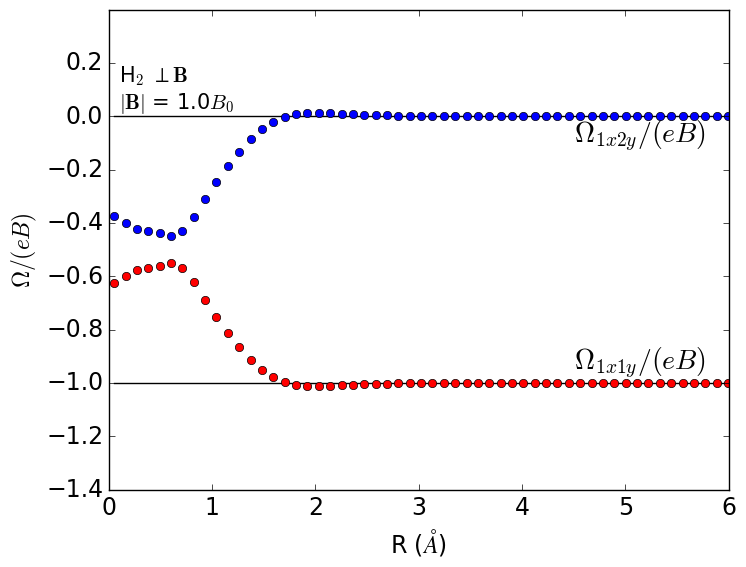} \\
\end{tabular}
\caption{Berry curvature of H$_2$ for the UHF $\beta\beta$ triplet state calculated with the cc-pVDZ+ basis set for orientations parallel (a $\&$ c) and perpendicular (b $\&$ d) to the magnetic field. The field is oriented along the $z$-axis with a strength of $|\theb| = 0.1B_0$ (a $\&$ b) and $|\theb| = 1.0B_0$ (c $\&$ d). Molecular orientation perpendicular to the field is along the $x$-axis. The cc-pVDZ+ basis is comprised of cc-pVDZ hydrogen basis functions centered at the nuclei and a cc-pVDZ Helium ghost center placed at the center of mass.}
\label{fig_uhf_dz-dz-dz_0.1_1.0}
\end{figure*}

Figures\,\ref{fig_uhf_sto3g-nobas-sto3g_0.1_1.0} and~\ref{fig_uhf_dz-dz-dz_0.1_1.0} show the Berry curvature of the $\mathrm{\beta}\beta$ UHF triplet state in the STO-3G and cc-pVDZ+ basis sets, respectively. In the parallel orientation, the values of $\Omega_{1x2y}$ tend asymptotically from the antiscreening regime toward zero as the bond distance is increased, while the values of $\Omega_{1x1y}$ tend asymptotically from the superscreening regime toward $-1eB_0$. In the perpendicular orientation, the elements of the Berry curvature stay within the bounds of 0 and $-1eB_0$ for the STO-3G basis set. In the cc-pVDZ+ basis, $\Omega_{1x2y}$ and $\Omega_{1x1y}$ move into the antiscreening or superscreening regimes, depending on the  field strength and bond distance. The same  features are present in both Figs.\,\ref{fig_uhf_sto3g-nobas-sto3g_0.1_1.0} and~\ref{fig_uhf_dz-dz-dz_0.1_1.0}, the main differences being the magnitudes of the curvature, which vary depending on the  field strength and basis set used.

Since the Berry curvature serves to screen the Lorentz force acting on the nuclei, it is noted that the elements $\Omega_{IxIy}$ and $\Omega_{IxJy}$ add up to the partial electronic charge associated with nucleus $I$. For H$_2$, this sum should be $-1$ for each nucleus, with the sum over both nuclei giving $-2$, indicating that the center of mass motion is completely screened as it should be in a neutral system. This is indeed what we observed in Figs.\,\ref{fig_sto3g-nobas-sto3g_0.1_1.0} -- \ref{fig_uhf_dz-dz-dz_0.1_1.0}. Depending on the basis set and method, the values of $\Omega_{IxIy}$ and $\Omega_{IxJy}$ change as a function of bond distance but their sum $-1$ is a constant.

\subsection{C$\mathbf{H}^{+}$ Molecule}

Berry-curvature values were calculated as a function of magnetic field strength in the range of $|\theb|$ = 0.0001$B_0$. to $|\theb| = B_0$ for the CH$^{+}$ molecule using finite difference as outlined in Section II. Calculations were performed for the RHF singlet state with the cc-pVDZ basis set at the zero field equilibrium geometry of 1.123\,$\text{\AA}$. The magnetic field was oriented along the $z$-axis, and calculations were performed both parallel and perpendicular to the magnetic field, with the perpendicular orientation of the molecule being specifically along the $x$-axis.

The Berry curvature tensor of CH$^+$ has a lower symmetry than that of H$_2$. 
For the parallel and perpendicular orientations of  CH$^{+}$ 
to the magnetic field, the structure of the Berry curvature tensor is
\begin{align}
    \boldsymbol \Omega &= \begin{pmatrix} \boldsymbol \Omega_\text{CC} & \boldsymbol \Omega_\text{CH} \\ \boldsymbol \Omega_\text{HC} & \boldsymbol \Omega_\text{HH} \end{pmatrix}
    \nonumber \\&=
    \left(\begin{array}{cccccc}
    0 & -\omega_\text{CC} & 0 & 0 & - \omega_\text{HC} & 0 \\ \omega_\text{CC} & 0 & 0 & \omega_\text{CH} & 0 & 0 \\
    0&0&0&0&0&0\\
    0& - \omega_\text{CH}& 0 & 0& -\omega_\text{HH} & 0 \\
    \omega_\text{HC} & 0 & 0 & \omega_\text{HH} & 0 & 0 \\
      0&0&0&0&0&0
    \end{array}\right)
\end{align}
obeying the overall symmetry $\boldsymbol \Omega_\text{CH} = - \boldsymbol \Omega_\text{HC}^{\mathrm{T}}$. However, since $\boldsymbol \Omega_\text{CH} \neq - \boldsymbol \Omega_\text{CH}^{\mathrm{T}}$ and $\boldsymbol \Omega_\text{HC} \neq - \boldsymbol \Omega_\text{HC}^{\mathrm{T}}$, the Berry force of CH$^+$  (unlike that of H$_2$) cannot be expressed in cross-product form. For interpretation purposes, we nevertheless calculate the partial charges of C and H from an average of $\omega_\mathrm{CH}/\omega_\mathrm{HC}$:
\begin{align}
q_\text{C}&= Q_\mathrm{CC} + Q_\mathrm{CH} = 
-\dfrac{\omega_\mathrm{CC}}{e B_z} -
\dfrac{\omega_\mathrm{CH} + \omega_\mathrm{HC}}{2e  B_z}\,,\\
q_\text{H}&= Q_\mathrm{HH} + Q_\mathrm{HC} = 
- \dfrac{\omega_\mathrm{HH} }{e B_z} -
\dfrac{\omega_\mathrm{HC} + \omega_\mathrm{CH}}{2e  B_z}\,. \end{align}

\begin{figure*}[h]
\centering
\begin{tabular}{ll}
(a) \\
\includegraphics[width=0.48\textwidth]{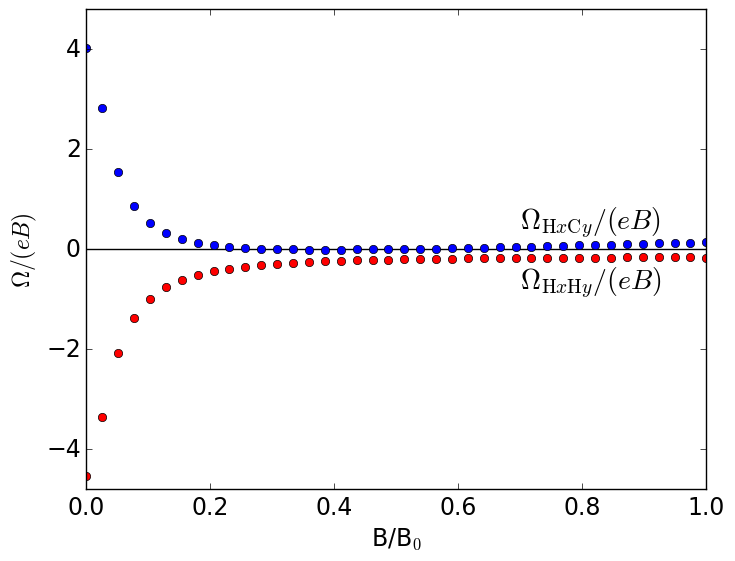} \\
(b) \\
\includegraphics[width=0.48\textwidth]{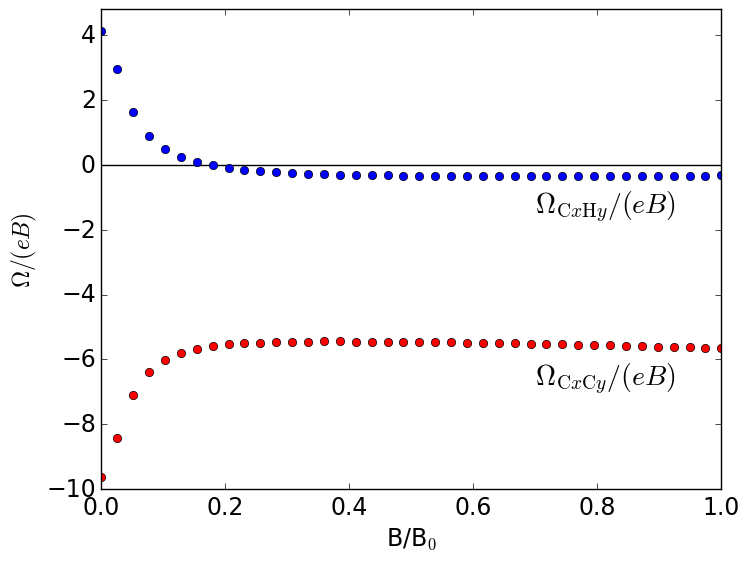} \\
\end{tabular}
\caption{Berry curvature of CH$^{+}$ for the RHF singlet state calculated with the cc-pVDZ basis set as a function of magnetic field strength. The magnetic field is oriented along the $z$-axis with the CH$^{+}$ molecule oriented perpendicular to the field along the $x$-axis. Panel (a) shows the HH and HC block elements, while panel (b) shows the CC and CH block elements. Bond distance is the zero-field equilibrium value of 1.123 $\text{\AA}$. Range of magnetic field strengths $|\theb| = 10^{-4}B_0$ to $|\theb|=B_0$.}
\label{fig_ch+_perp_h1h2_c1c2}
\end{figure*}

\begin{figure*}[h]
\centering
\begin{tabular}{ll}
(a) \\
\includegraphics[width=0.48\textwidth]{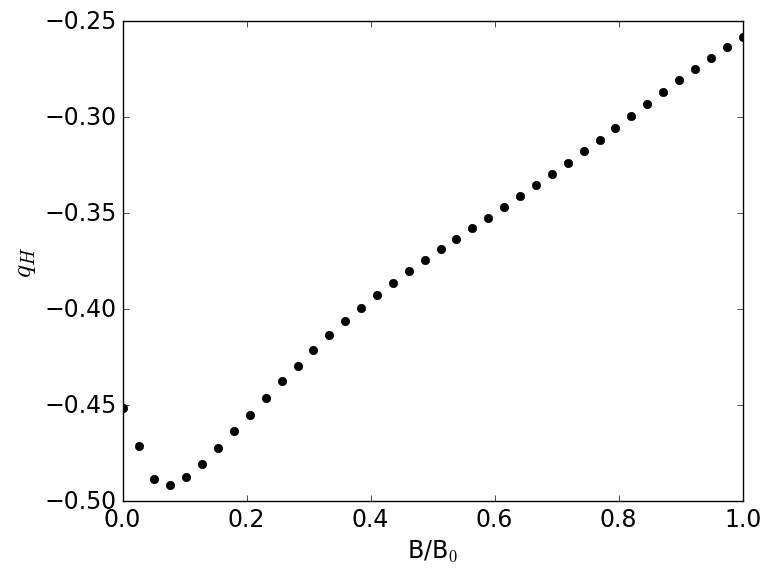} \\
(b) \\
\includegraphics[width=0.48\textwidth]{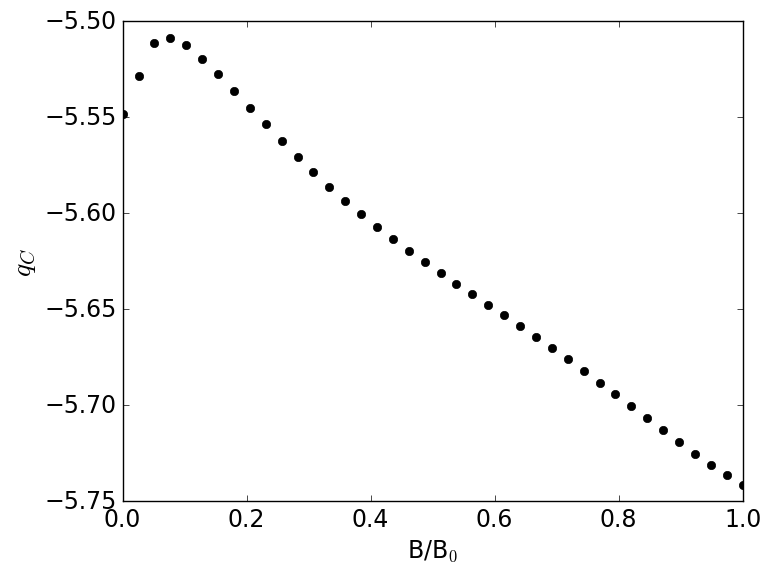} \\
\end{tabular}
\caption{Partial electronic charge for hydrogen $q_\text H$ (a) and carbon $q_\text C$ (b) calculated from Berry curvature elements for the RHF singlet state with the cc-pVDZ basis set as a function of magnetic field strength. The magnetic field is oriented along the $z$-axis with the CH$^{+}$ molecule oriented perpendicular to the field along the $x$-axis. Bond distance is the zero field equilibrium value of 1.123 $\text{\AA}$. Range of magnetic field strength is $|\theb| = 10^{-4}B_0$ to $|\theb| = B_0$.}
\label{fig_ch+_perp_hsum_csum}
\end{figure*}

\begin{figure*}[h]
\centering
\begin{tabular}{ll}
(a) & (b) \\
\includegraphics[width=0.48\textwidth]{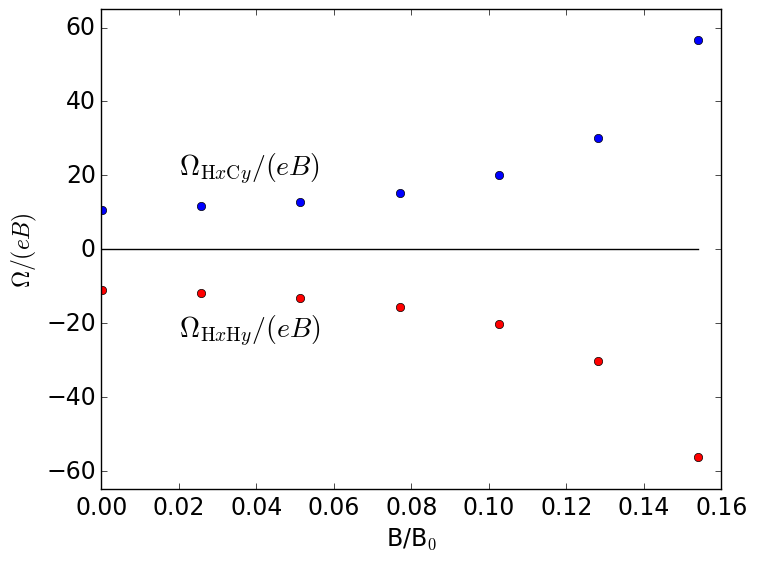} &
\includegraphics[width=0.48\textwidth]{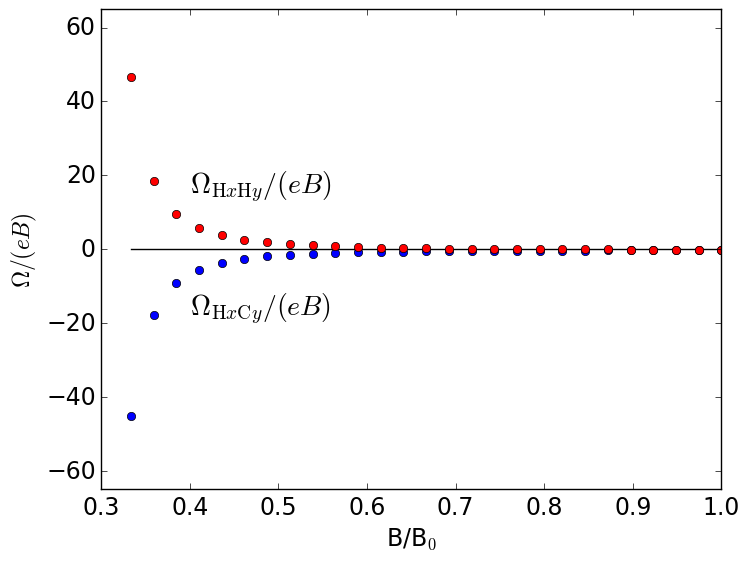} \\
(c) & (d) \\
\includegraphics[width=0.48\textwidth]{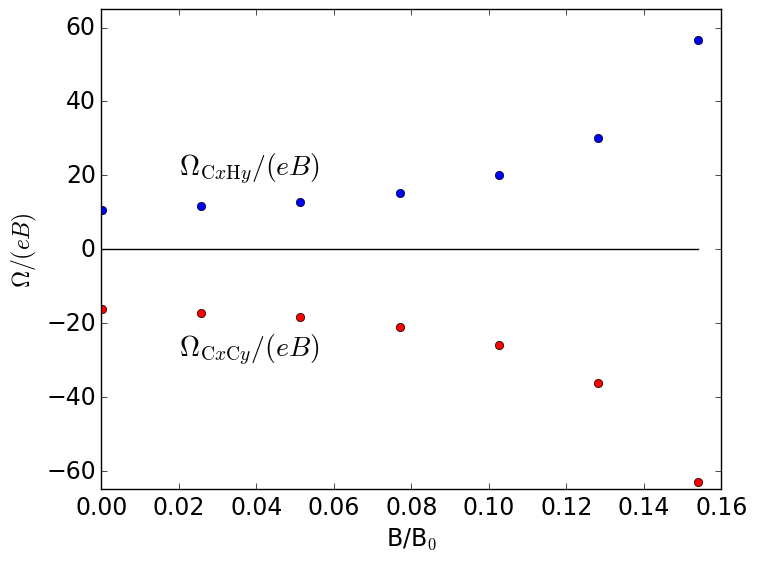} &
\includegraphics[width=0.48\textwidth]{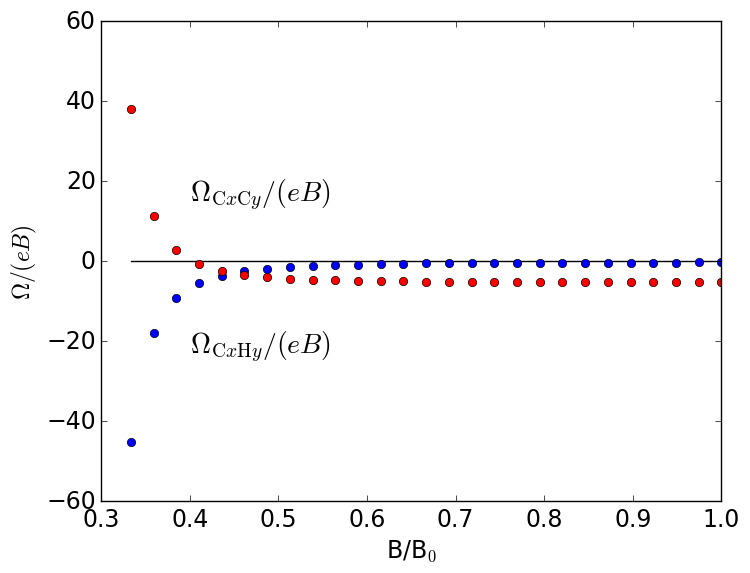} \\
\end{tabular}
\caption{Berry curvature of CH$^{+}$ for the RHF singlet state calculated with the cc-pVDZ basis set as a function of magnetic field strength. The magnetic field is oriented along the $z$-axis with the CH$^{+}$ molecule oriented parallel to the field along the $z$-axis. Panels (a) and (b) show the HH and HC block elements, while panels (c) and (d) show the CC and CH block elements. A level crossing occurs between $0.2B_0$ and $0.3B_0$, so data from this region was omitted in the plots. Bond distance was the zero field equilibrium value of 1.123 $\text{\AA}$. Range of magnetic field strength was $|\theb|= 10^{-4}B_0$ to $|\theb| = B_0$.}
\label{fig_ch+_parallel_h1h2_c1c2}
\end{figure*}

\begin{figure*}[h]
\centering
\begin{tabular}{ll}
(a) & (b) \\
\includegraphics[width=0.48\textwidth]{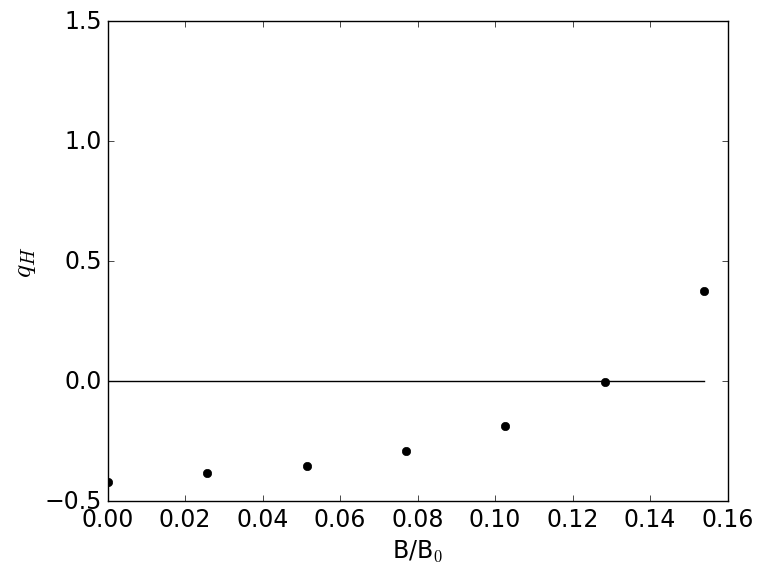} &
\includegraphics[width=0.48\textwidth]{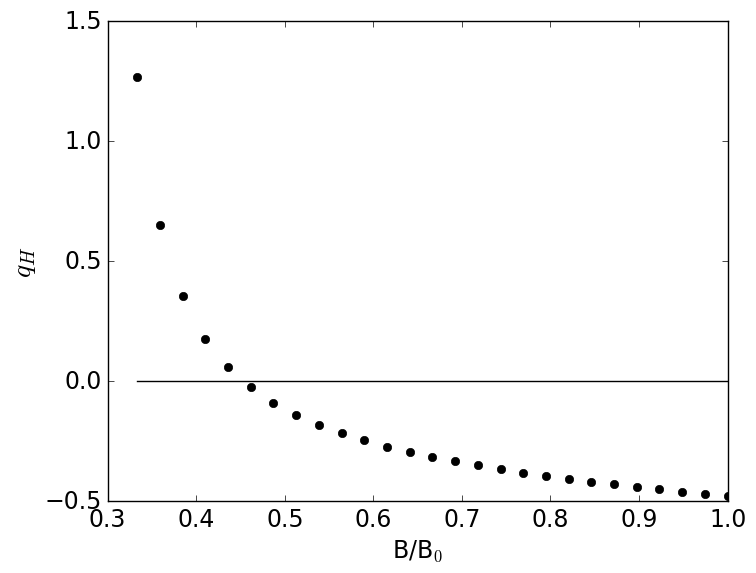} \\
(c) & (d) \\
\includegraphics[width=0.48\textwidth]{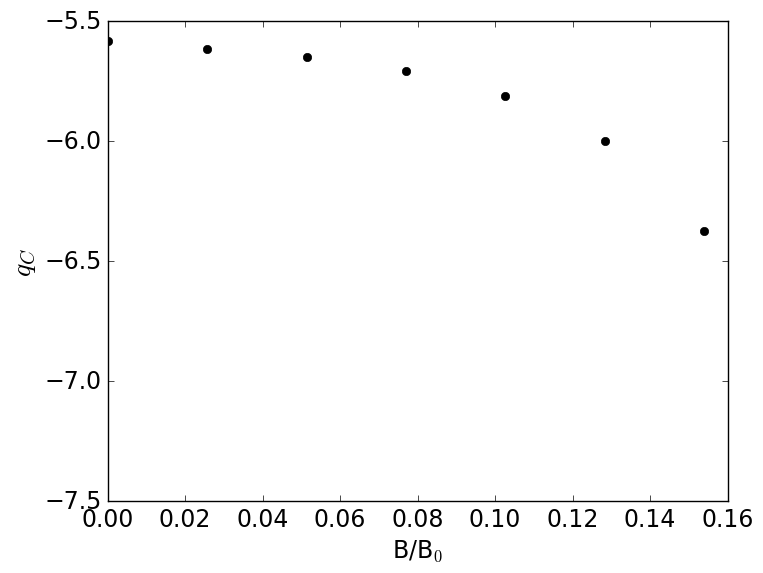} &
\includegraphics[width=0.48\textwidth]{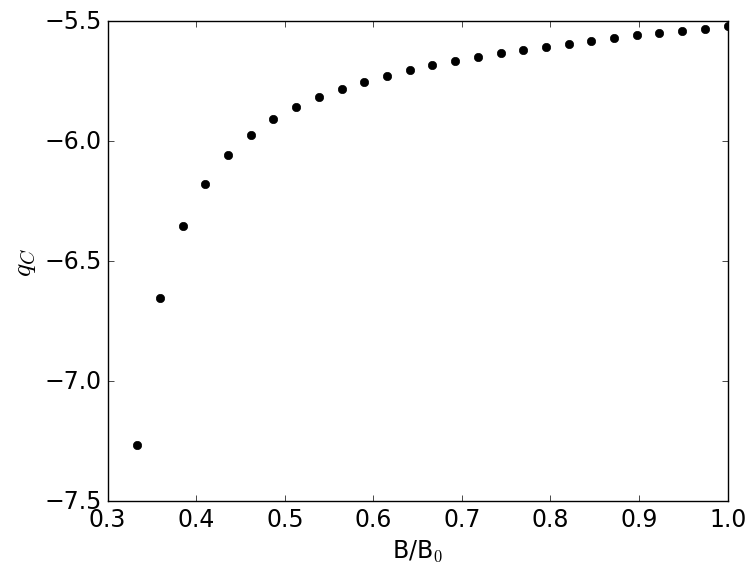} \\
\end{tabular}
\caption{Partial electronic charge for hydrogen $q_\text H$ (a $\&$ b) and carbon $q_\text C$ (c $\&$ d) calculated from Berry curvature elements for the RHF singlet state with the cc-pVDZ basis set as a function of magnetic field strength. The magnetic field is oriented along the $z$-axis with the CH$^{+}$ molecule oriented parallel to the field along the $z$-axis. A level crossing occurs in the region of $|\theb| = 0.2B_0$ to $|\theb| = 0.3B_0$, so data from this region was omitted in the plots. Bond distance is the zero field equilibrium value of 1.123 $\text{\AA}$. Range of magnetic field strength is $|\theb| = 10^{-4}B_0$ to $|\theb| = B_0$.}
\label{fig_ch+_parallel_hsum_csum}
\end{figure*}

For the perpendicular orientation of CH$^{+}$ to the magnetic field, 
Fig.\,\ref{fig_ch+_perp_h1h2_c1c2} shows the values of the individual $xy$ components for the hydrogen, carbon, and mixed blocks of the Berry curvature tensor plotted as a function of magnetic field strength. For each atom, the partial charge as a function of  field strength is plotted in Fig.\,\ref{fig_ch+_perp_hsum_csum}. The corresponding data for the parallel
orientation are plotted in Figs.\,\ref{fig_ch+_parallel_h1h2_c1c2} and \ref{fig_ch+_parallel_hsum_csum}.

In the perpendicular case, it is seen from Fig.\,\ref{fig_ch+_perp_hsum_csum}  that the charge is distributed unequally; moreover, the polarity of this distribution is enhanced as the field is increased. The electronic partial charge associated with the hydrogen nucleus tends toward zero as the field strength increases to $1.0B_0$, while the  partial charge associated with the carbon nucleus tends toward $-6$. The magnitude of the sum of the two partial charges is always equal to the total number of electrons.

In the parallel case, a level crossing is observed in the region of magnetic field strength of 0.2$B_0$--0.3$B_0$. As such, the values of the Berry curvature tensor in this region have been excluded from Figs.\,\ref{fig_ch+_parallel_h1h2_c1c2} and \ref{fig_ch+_parallel_hsum_csum}. The values of the partial charges given in Fig.\,\ref{fig_ch+_parallel_hsum_csum} increase for hydrogen and decrease for carbon in the vicinity of the level crossing, but the values in the limit of weak field and strong field are very similar, in contrast with the perpendicular orientation.

\section{Conclusions}

In this work, we have presented a general scheme for calculating the Berry curvature by finite differences. This was accomplished through a  phase correction of each perturbed wave function in the finite-difference procedure that results in a vanishing geometric vector potential. In the context of finite difference, such a correction is equivalent to the calculation of the derivatives in a consistent gauge. The scheme has been implemented in the program package {\sc london},\cite{LondonProgram} which uses London atomic orbitals for gauge-origin invariant calculations of molecules in a magnetic field. It should be noted that, while we have reported on Hartree--Fock wave functions, the method is generally applicable and can be used in the cases of DFT and FCI, for example.

Basis-set studies on small atoms demonstrated
that the reliability of the calculated Berry curvature depends critically on the use of London atomic orbitals---without London orbitals, the results are unreliable and basis-set convergence is slow; with London orbitals, convergence is rapid and a qualitatively correct Berry curvature is obtained even in a minimal basis.

The H$_2$ and CH$^{+}$ molecules were studied using London orbitals. The Berry curvature was calculated as a function of bond distance for H$_2$ at both the RHF and UHF levels of theory and as a function of magnetic field strength for CH$^{+}$ at the RHF level of theory. For short
bond distances, less than 0.5\,bohr, the behavior of the  Berry curvature was found to be very sensitive to the basis set used.
The H$_2$ results at the RHF/STO-3G level of theory agree with previous calculations of the Berry curvature for H$_2$ in a minimal London Slater basis.

Trends of the Berry curvature across multiple basis sets and  field strengths were investigated. Berry antiscreening and superscreening  were observed across multiple basis sets, indicating that these are real physical phenomena and not solely basis-set effects. For  CH$^{+}$  in the perpendicular 
orientation, the polarity of the electronic partial charges on hydrogen and carbon increase with increasing field strength. In the parallel orientation, by contrast, the partial charges are roughly the same in the weak- and strong-field limits, with a level crossing occuring at intermediate field strengths.

Calculation of the Berry curvature from finite difference opens up the possibility of accurately calculating the screening force due to the electrons in \textit{ab initio} molecular dynamics simulations. The screening of the Lorentz force is of crucial importance and can change the dynamics of atoms and molecules in a magnetic field qualitatively. Consequently, calculation of the Berry curvature will be important for future work pertaining to molecular dynamics in magnetic fields.  

\section*{Supplementary Material}

See Supporting Information for Berry curvature plots with the cc-pVTZ basis set.

\section*{Acknowledgements}

We thank Laurenz Monzel, Ansgar Pausch, and Wim Klopper (Karlsruher Institut f\"ur Technologie) for helpful discussions. This work was supported by the Research Council of Norway through ‘‘Magnetic Chemistry’’ Grant No.\,287950 and CoE Hylleraas Centre for Quantum Molecular Sciences Grant No.\,262695.

\section*{References}


\end{document}